A Large Scale Magneto-ionic Fluctuation in the Local Environment of Periodic Fast Radio Burst Source, FRB 20180916B

R. Mckinven ,[1,2,3,4] B.M. Gaensler ,[3,4] D. Michilli ,[5,6] K. Masui ,[6,5] V.M. Kaspi ,[1,2] M. Bhardwaj ,[1,2]
T. Cassanelli ,[4,3] P. Chawla ,[7] F. (Adam) Dong ,[8] E. Fonseca ,[9,10,11] C. Leung ,[6,5] D.Z. Li ,[12] C. Ng ,[3]
C. Patel ,[1,3] E. Petroff ,[1,2,7] A.B. Pearlman ,[1,2] Z. Pleunis ,[3] M. Rafiei-Ravandi ,[1,2] M. Rahman,[13]
K.R. Sand ,[1,2] K. Shin ,[6,5] P. Scholz ,[3] I. H. Stairs ,[8] K. Smith ,[14] J. Su ,[1] and S. Tendulkar [15,16]

[1]Department of Physics, McGill University, 3600 rue University, Montréal, QC H3A 2T8, Canada
[2]McGill Space Institute, McGill University, 3550 rue University, Montréal, QC H3A 2A7, Canada
[3]Dunlap Institute for Astronomy & Astrophysics, University of Toronto, 50 St. George Street, Toronto, ON M5S 3H4, Canada
[4]David A. Dunlap Department of Astronomy & Astrophysics, University of Toronto, 50 St. George Street, Toronto, ON M5S 3H4, Canada
[5]Department of Physics, Massachusetts Institute of Technology, 77 Massachusetts Ave, Cambridge, MA 02139, USA
[6]MIT Kavli Institute for Astrophysics and Space Research, Massachusetts Institute of Technology, 77 Massachusetts Ave, Cambridge, MA 02139, USA
[7]Anton Pannekoek Institute for Astronomy, University of Amsterdam, Science Park 904, 1098 XH Amsterdam, The Netherlands
[8]Dept. of Physics and Astronomy, 6224 Agricultural Road, Vancouver, BC V6T 1Z1 Canada
[9]Lane Department of Computer Science and Electrical Engineering, 1220 Evansdale Drive, PO Box 6109 Morgantown, WV 26506, USA
[10]Center for Gravitational Waves and Cosmology, West Virginia University, Chestnut Ridge Research Building, Morgantown, WV 26505, USA
[11]Department of Physics and Astronomy, West Virginia University, P.O. Box 6315, Morgantown, WV 26506, USA
[12]Cahill Center for Astronomy and Astrophysics, California Institute of Technology, 1216 E California Boulevard, Pasadena, CA 91125, USA
[13]Sidrat Research, PO Box 73527 RPO Wychwood, Toronto, Ontario, M6C 4A7, Canada
[14]Perimeter Institute for Theoretical Physics, 31 Caroline Street N, Waterloo ON N2L 2Y5 Canada
[15]Department of Astronomy and Astrophysics, Tata Institute of Fundamental Research, Mumbai, 400005, India
[16]National Centre for Radio Astrophysics, Post Bag 3, Ganeshkhind, Pune, 411007, India

## ABSTRACT

Fast radio burst (FRB) source 20180916B exhibits a 16.33-day periodicity in its burst activity. It is as of yet unclear what proposed mechanism produces the activity, but polarization information is a key diagnostic. Here, we report on the polarization properties of 44 bursts from FRB 20180916B detected between 2018 December and 2021 December by CHIME/FRB, the FRB project on the Canadian Hydrogen Intensity Mapping Experiment the Canadian Hydrogen Intensity Mapping Experiment. In contrast to previous observations, we find significant variations in the Faraday rotation measure (RM) of FRB 20180916B. Over the nine month period 2021 April−2021 December we observe an apparent secular increase in RM of $\sim 50$ rad m$^{-2}$ (a fractional change of over 40%) that is accompanied by a possible drift of the emitting band to lower frequencies. This interval displays very little variation in the dispersion measure ($\Delta$DM $\lesssim 0.8$ pc cm$^{-3}$) which indicates that the observed RM evolution is likely produced from coherent changes in the Faraday-active medium's magnetic field. Burst-to-burst RM variations appear unrelated to the activity cycle phase. The degree of linear polarization of our burst sample ($\gtrsim 80\%$) is consistent with the negligible depolarization expected for this source in the 400-800 MHz bandpass of CHIME. FRB 20180916B joins other repeating FRBs in displaying substantial RM variations between bursts. This is consistent with the notion that repeater progenitors may be associated with young stellar populations by their preferential occupation of dynamic magnetized environments commonly found in supernova remnants, pulsar wind nebulae or near high mass stellar companions.

## 1. INTRODUCTION

Fast radio bursts (FRBs; Lorimer et al. 2007) are short (microsecond to millisecond) bursts of radio emission that have been observed over a wide range of frequencies and are characterized by a dispersed signal that usually indicates source locations at cosmological distances (see Cordes & Chatterjee 2019; Petroff et al. 2019, 2021, for reviews of the phenomenon). With the release of a catalog containing over 500 fast radio burst (FRB) detections, the FRB survey operating on the Canadian Hydrogen Intensity Mapping Experiment (CHIME/FRB; CHIME/FRB Collaboration et al. 2018) has provided an unprecedented opportunity to study these sources from a variety of perspectives (CHIME/FRB Collaboration et al. 2021). This catalog has inspired several

Corresponding author: R. Mckinven
ryan.mckinven@mcgill.ca

recent studies that leverage different properties of the sample to constrain the FRB population (e.g., Pleunis et al. 2021a; Josephy et al. 2021; Rafiei-Ravandi et al. 2021; Chawla et al. 2021) as well as the identification of individual sources favorable for follow-up observations. Noticeably absent from these analyses until very recently were systematic studies of the polarization properties of FRBs.

Our current understanding of FRB polarization is mainly determined by polarimetric observations from a handful of prolific repeating sources. Similar to one-off FRBs, the polarized signal of repeat bursts are characterized by a few well-known quantities: the linear and circular polarized fraction ($L/I$ & $V/I$), the Faraday rotation measure (RM) and the position angle (PA). For each of these quantities, both the average value as well as any evolution over the burst duration provide information on the nature of the source and the intervening, magnetized plasma. However, only repeaters enable these quantities to be tracked over long intervals of time through polarimetric monitoring of repeat bursts. This feature has proved to be a powerful tool, with multi-epoch observations of several repeaters demonstrating substantial RM evolution that suggests a preferential occupation of repeating sources in dynamic magneto-ionic environments. Many of these observations display significant RM variations on intra-day (e.g., Luo et al. 2020; Hilmarsson et al. 2021a), multi-week (e.g., Xu et al. 2021) and/or multi-year (Hilmarsson et al. 2021b) timescales. These RM variations are very often irregular and discrepant with the secular decrease expected if the RM is dominated by a shocked, uniform medium (e.g. Piro & Gaensler 2018) and instead suggest a high degree of non-uniformity in the local environment of the source.

The preferential occupation of repeaters in dynamic environments appears to be independently supported by recent multiband observations of five sources (FRB 20121102A, FRB 20190520B, FRB 20190303A, FRB 20190417A, FRB 20201124A), all of which showed a clear trend of a decreasing degree of linear polarization at lower frequencies (Feng et al. 2022). This trend is well characterized by a single parameter, the rotation-measure-scatter ($\sigma_{RM}$), which quantifies the degree of stochastic Faraday rotation encountered by scattered FRB emission. The possible correlation $\sigma_{RM}$ with temporal scattering, $\tau_s$ ($\tau_s \propto \sigma_{RM}^{1.0 \pm 0.2}$; Feng et al. 2022), suggests that temporal scattering and RM scatter originate from the same region. Furthermore, the repeaters exhibiting the most extreme $\sigma_{RM}$ values, FRB 20121102A and FRB 20190520B, are those associated with a compact persistent radio source (PRS; Marcote et al. 2017; Niu et al. 2021). Yang et al. (2022) argue that this observation is consistent with expectations if the PRS is a supernova remnant or pulsar wind nebula and make predictions for the $\tau_s \propto \sigma_{RM}^{\alpha}$ relation under different scenarios of the Faraday-active plasma. Whether all repeaters can be described under a similar frame-work, even for repeaters without any apparent PRS association, remains an open question. Given the apparent sensitivity of the polarized signal to the local environments of their sources, polarimetric monitoring of repeaters should be an incredibly useful tool for discriminating among the various FRB progenitor/emission models.

In the case of FRB 20180916B, polarimetric observations are particularly motivated given the periodic bursting activity displayed from this source (CHIME/FRB Collaboration et al. 2020a), most recently measured to be $16.33 \pm 0.12$ days (Pleunis et al. 2021b) Some models put forth to explain this periodicity make explicit predictions on the evolution of certain polarization properties. For example, models invoking a slowly rotating or precessing FRB-emitting magnetar predict subtle PA variations across bursts that are correlated to the activity cycle (Li & Zanazzi 2021). Conversely, models explaining the periodicity in terms of a binary orbit with a stellar companion suggest the possibility of modulation in the RM if the eclipsing wind of the companion contributes to the observed Faraday rotation (Lyutikov et al. 2020; Ioka & Zhang 2020).

Despite the relevance of polarization information for FRB 20180916B, the published sample of polarimetric observations for this source remains quite limited. First detected by CHIME/FRB with RM $= -114.6 \pm 0.6$ rad m$^{-2}$ (CHIME/FRB Collaboration et al. 2019), subsequent detections of FRB 20180916B at several other frequency bands between 100 MHz and 1.7 GHz (Chawla et al. 2020; Pilia et al. 2020; Marcote et al. 2020; Pleunis et al. 2021b; Nimmo et al. 2021a; Sand et al. 2021) have established fairly stable RM behavior over time. However, recent observations with LOFAR over $110 - 188$ MHz demonstrated small but significant $\sim 2 - 3$ rad m$^{-2}$ ($\sim 2\%$ fractional) RM variations and a systematically lower fractional polarization at lower frequencies that is attributed to depolarization effect of stochastic Faraday rotation produced by scattered emission (Pleunis et al. 2021b; Feng et al. 2022).

This paper reports on polarimetric observations of a sample of 44 bursts from FRB 20180916B. This sample greatly expands on previous polarized observations of this source, enabling polarized quantities such as RM and $L/I$ to be tracked over multi-year timescales and to be correlated with other quantities including activity cycle phase. The paper is organized as follows: Section 2 reviews the observations and analysis methods, Section 3 summarizes results and is followed by Section 4 which explores the implications of these results for different theoretical models and how these results compare to observations of other FRB sources. Concluding remarks are provided in Section 5.

## 2. METHODS

### 2.1. *Data Reduction and Observations*

As reviewed by Michilli et al. (2021), CHIME/FRB has a baseband backend that enables the complex channelized voltages of each correlator input to be recorded upon successful trigger of the real-time FRB search pipeline (CHIME/FRB Collaboration et al. 2018). The two polarizations of each of the 1024 CHIME dual-linear feeds are recorded separately, enabling polarimetric analysis that is otherwise unavailable with the real-time intensity data. Data are reduced by first beamforming on the best sky position of the source and converting from complex voltages $(X, Y)$ to real valued Stokes parameters $(I, Q, U, V)$ through the transformation[1],

$$
\begin{aligned}
I &= \langle |X|^2 + |Y|^2 \rangle \\
Q &= \langle |X|^2 - |Y|^2 \rangle \\
U &= \langle 2\, real(XY^*) \rangle \\
V &= \langle -2\, imag(XY^*) \rangle.
\end{aligned}
\tag{1}
$$

Here, Stokes $I, Q, U, V$ is the canonical way of representing the signal such that Stokes $I$ refers to the total intensity of the emission, Stokes $Q$ and $U$ correspond to the linearly polarized component and Stokes $V$ refers to the circularly polarized component. These data are natively channelized with a frequency and time resolution of 390 kHz and 2.56 $\mu$s, respectively. Further information on the specifics of the data reduction and analysis are provided by Mckinven et al. (2021), who describe an automated polarization pipeline for CHIME/FRB.

Observations reported in this paper (see Table 1) correspond to events that successfully triggered baseband callback data over a nearly three year period between 2018 December and 2021 December. Baseband data were successfully recorded for 44 bursts. Due to resource limitations and the threat posed by false positives, a higher S/N threshold has been historically set for callbacks of baseband versus intensity data. Hence, the detections reported here are not entirely reflective of the total number of detections, which will be reported elsewhere. The baseband callback system, described in detail by Michilli et al. (2021), corresponds to a memory buffer capable of recording up to $\sim 20$ seconds of complex voltage data from each of the 1024 dual-linear feeds. This memory limit corresponds to an upper limit of $\mathrm{DM} \sim 1000\,\mathrm{pc\,cm^{-3}}$, for full baseband callbacks. Bursts with DMs greater than this have dispersive delays that exceed time limits of the buffer and result in data being lost at the top of the CHIME band. Fortunately, the DM of FRB 20180916B is well below this this limit, ensuring that no data are lost.

### 2.2. Data Analysis

#### 2.2.1. Dispersion Measures

Two varieties of dispersion measures (DMs) were determined for this paper, $\mathrm{DM_{S/N}}$ and $\mathrm{DM_{struct}}$. $\mathrm{DM_{S/N}}$ corresponds to the DM at which the band-averaged S/N reaches a maximum peak value, while $\mathrm{DM_{struct}}$, in effect, optimizes for the alignment of frequency-time substructure. In both cases, coherent dedispersion to a nominal DM value is applied prior to DM optimization. DM optimization is performed using downsampled data. For each burst a downsampling factor, $\mathrm{n_{down}}$ (see Table 1), is determined by iteratively sampling from $2^N$ ($N = 1, 2, 3, ...$) until the peak S/N of the burst exceeds 8. For our sample, $\mathrm{n_{down}}$ ranges from 1 to 512. This corresponds to a time resolution spanning $2.56\,\mu s - 1.31$ ms.

For $\mathrm{DM_{S/N}}$, we incoherently dedisperse to a set of trial DMs and determine the peak S/N of the band-averaged profile at each trial. The resulting S/N measurements as a function of DM are then fit with a composite two-Gaussian function, with the peak location corresponding to $\mathrm{DM_{S/N}}$ and the width at which it falls by 1 S/N unit, its uncertainty. A two-Gaussian function was used since the S/N of many bursts were found to be sensitive to small changes in DM, which could not be adequately captured by a single Gaussian.

Many bursts from the sample studied here display complex morphology, replicating the down-drifting structure commonly seen from repeating sources (e.g., Hessels et al. 2019). For such bursts, $\mathrm{DM_{S/N}}$ is vulnerable to confusing burst structure with additional dispersion and results in DM measurements that are systematically overestimated. We find this to be the case for the sample studied here and, therefore, do not report $\mathrm{DM_{S/N}}$ values to avoid confusion. This issue is circumvented with methods that optimize vertical alignment of bursts' substructure. Here, we follow previous CHIME/FRB papers (e.g., CHIME/FRB Collaboration et al. 2019; Fonseca et al. 2020; CHIME/FRB Collaboration et al. 2020b), employing the $\mathtt{DM\_phase}$ package[2] (Seymour et al. 2019) to calculate the phase coherence of emission in all frequency channels and find the optimal value for $\mathrm{DM_{struct}}$. Measurements of each burst's RM are determined using data from each burst dedispersed to its corresponding $\mathrm{DM_{struct}}$ value.

#### 2.2.2. Rotation Measures

The polarization pipeline operating on CHIME/FRB data employs two separate methods for measuring the RMs of polarized sources: a non-parametric method referred to as *RM-synthesis* and a parametric method known as *QU-fitting*. To first order, FRB emission is in the *Faraday thin* regime where

---

all of the linear polarized signal is Faraday rotated by the same amount. In this regime, RM-synthesis and QU-fitting are roughly equivalent, however, RMs determined here from QU-fitting are considered more accurate than those obtained from RM-synthesis thanks to a refined model fit that allows simultaneous characterization of dominant systematics (see Section 2.2.3). A full description of these two methods and their implementation within the CHIME/FRB polarization pipeline is given by Mckinven et al. (2021). Here, we provide a brief summary.

*RM-synthesis:*

RM-synthesis (e.g., Brentjens & de Bruyn 2005) is a popular technique for measuring Faraday rotation and is suitable in cases where the RM is not a priori known. The technique amounts to a Fourier-like transformation, such that,

$$F(\phi) = \int_{-\infty}^{\infty} P(\lambda^2) e^{-2i\phi\lambda^2} d\lambda^2, \qquad (2)$$

measures the linear polarized intensity at different Faraday depths, $\phi$, or RMs. Here, the complex quantity $P(\lambda^2) = Q(\lambda^2) + iU(\lambda^2)$ represents the linear polarization as a function of the wavelength dependence expected from Faraday rotation (i.e., $\lambda^2$). The resulting function, $F(\phi)$, referred to as the *Faraday Dispersion Function* (FDF), can be used to measure the RM as well as the degree to which polarized emission is distributed among multiple $\phi$ or RM values. FRB emission occupies the "Faraday thin" regime when, to first order, all emission is Faraday rotated by the same amount per frequency. In this regime, $\mathrm{RM_{FDF}}$ corresponds to the Faraday depth at which $\mathbf{F}(\phi)$ peaks with the uncertainty, $\sigma_{\mathrm{RM_{FDF}}}$, determined by the full-width-half-max (FWHM) and S/N of the FDF peak such that, $\sigma_{\mathrm{RM_{FDF}}} = \frac{\mathrm{FWHM}}{2\,\mathrm{S/N}}$. The resulting FDF can be cleaned of artefacts introduced by the limited bandpass of the observations by applying a deconvolution algorithm presented by Heald (2009). Cleaned FDFs for the bursts studied here can be found in Appendix B.

*QU-fitting:*

An alternative method for determining an RM is to invoke a model that adequately fits the oscillations in Stokes $Q$ and $U$ as a function of $\lambda^2$ introduced by Faraday rotation (O'Sullivan et al. 2012). This method enjoys greater flexibility, enabling parameters to be added that describe additional properties, including those introduced by the instrument. Models that simultaneously capture the polarized astrophysical and instrumental signal are implemented into a Nested Sampling QU-fitting framework (Mckinven et al. 2021). The default mode of the pipeline is to fit for the astrophysical parameters of $p$ (linear polarized fraction), $\mathrm{RM_{QU}}$, $\psi_0$ (polarization position angle at infinite frequency), and $\tau$

(a physical delay between the two linear polarizations). The precise formulation of the model and the likelihood function through which parameter estimation occurs can be found by consulting Mckinven et al. (2021). Polarized spectra and their best-fitting $QU$ models for the sample studied here can be found in Appendix B.

### 2.2.3. *Effects of a Non-zero Physical Delay*

A non-zero $\tau$ value can result in leakage between Stokes $U$ and $V$. In principle, any attempt to measure RM in the presence of a non-negligible $\tau$ must account for an intrinsic circularly polarized signal. While in theory, bursts displaying $|V|/I \sim 100\%$ are possible, in practice, circular polarization from FRBs is often found to be consistent with $0\%$. We therefore simplify our QU-fit model by assuming a negligible astrophysical Stokes $V$. The quality of our resulting fits seems to justify this choice and we find no evidence for bursts with fit residuals that would indicate substantial circular polarization such as is the case for some bursts from FRB 20201124A (Kumar et al. 2021; Xu et al. 2021). That said, a marginal Stokes V signal seen from other FRB sources (e.g. Petroff et al. 2015; Caleb et al. 2018) may be present in these observations, but is beyond our ability to confidently characterize given our current understanding of our systematics. For example, cross polarization between linear feeds likely represents a substantial contribution to the residual Stokes $V$. As a consequence, any observed Stokes $V$ (e.g., burst 3 in Figure 1) is assumed to be a result of secondary sources of leakage and we do not report the circular polarization fraction $(V/I)$ in this paper.

We do not report individual best-fit $\tau$ or $\psi_0$ values since we consider both these quantities to be under-constrained. In the case of $\tau$, values on the order of 1 ns are regularly fit from the data but are accompanied by degeneracies with RM and $\psi_0$ in which multiple values of ns can produce convincing fits. This degeneracy is more commonly seen in narrowband bursts where there is insufficient bandwidth coverage to distinguish among the possibilities. Errors on $\tau$ can affect best-fit measurements of $\psi_0$ and RM. In the case of $\psi_0$, a non-zero $\tau$ adds a differential phase between the two linear polarizations that leaves Stokes U, but not Stokes Q, affected. This, combined with the absence of robust polarimetric calibration[3], imply a high degree of uncertainty in $\psi_0$ values and we do not calculate them for this paper. Meanwhile, a non-zero $\tau$ often leads to an ambiguity in the sign of the RM if unaccounted for, as previously demonstrated by Mckinven et al. (2021). This ambiguity appears to be resolved here with our QU-fitting implementation capturing the $U - V$ leakage using a fitted $\tau$ parameter that consistently results in a S/N boost in the linear polarized component and greater precision in the RM measurement. This is one of the main advantages of QU-fitting, and thus $RM_{QU}$ should be considered a more reliable estimate of the true RM over $RM_{FDF}$[4].

---

[3] Polarimetric calibration remains a challenge for transiting telescopes such as CHIME.

### 2.2.4. *PA Curves*

The PA behavior over the burst duration (PA curves) displayed in Figures 1 are calculated from the subband averaged Stokes $Q$ and $U$ profiles corrected for Faraday rotation, $Q_{derot}$ and $U_{derot}$,

$$PA = \frac{1}{2} \tan^{-1} \left( \frac{U_{derot}(t)}{Q_{derot}(t)} \right) \quad rad. \quad (3)$$

Here, $Q_{derot}$ and $U_{derot}$ have been *de-rotated* by assuming RM and $\psi_0$ parameters obtained from QU-fitting and applying Equation 5 of Mckinven et al. (2021). Here, $\psi_0$ corresponds to the polarization angle at infinite frequency, which corresponds to the PA in fully calibrated data. In the case of CHIME, this level of calibration is not possible due to residual systematics and the fitted value of $\psi_0$ is only used to center PA curves near $\sim 0$ degrees. PA curves displayed here are therefore only intended to characterize the relative PA evolution *within* bursts and not the precise value of the PA or its evolution *across* bursts. Uncertainties on PA measurements are estimated through conventional error propagation (e.g., see Equation A3 from Vernstrom et al. 2019). PA values are shown only where the linear polarization S/N $\geq 3$. As such, a fraction of these PA data points will correspond to noise and will appear to have underestimated error bars due to our selected S/N threshold.

### 2.3. *Galactic Rotation and Dispersion Measure Estimates*

The Galactic RM contribution and associated uncertainties are estimated from the updated all-sky Faraday Sky map of Hutschenreuter et al. (2021). This map reconstructs the Galactic RM contribution, $RM_{MW}$, using RM measurements of polarized extragalactic sources. Using the refined position of this source provided by the interferometric localization with the European VLBI Network (EVN; Marcote et al. 2020), we estimate the Galactic RM contribution at the position of FRB 20180916B to be $RM_{MW} = -94 \pm 45$ rad m$^{-2}$. Meanwhile, the Galactic DM contribution is estimated using the PyGEDM package[5], a python interface to the YMW16 (Yao et al. 2017) and NE2001 (Cordes & Lazio 2002, 2003) electron density models. These two models yield Galactic DM estimates of $DM_{MW}^{ymw} \sim 325$ pc cm$^{-3}$ and $DM_{MW}^{ne2001} \sim 199$ pc cm$^{-3}$, respectively. These estimates do not include the DM contribution of the Galactic halo which is expected to have an average value of $\sim 30 - 50$ rad m$^{-2}$, depending on the halo model assumed (e.g., Dolag et al. 2015; Yamasaki & Totani 2020).

### 2.4. *Ionospheric Rotation Measure Subtraction*

Earth's ionosphere imparts a small but measurable contribution to the observed Faraday rotation of any linear polarized source. This contribution, $RM_{iono}$, must be corrected to determine the significance of any RM variations. $RM_{iono}$ can vary significantly depending on time of day, solar cycle and pointing (Mevius 2018a). However, for a transit telescopes, detections preferentially occur where sensitivity is greatest. In the case of CHIME, this is along the major (N-S) axis of the hyperbolic dishes (CHIME/FRB Collaboration et al. 2018). As such, most FRB detections occur very near (a few degrees) of zenith. This greatly reduces the variability introduced by $RM_{iono}$ that would otherwise be obtained if the ionosphere were more uniformly sampled. This appears to be validated by recent pulsar RMs from CHIME/Pulsar observations, which found general agreement with reference RM values despite not accounting for $RM_{iono}$ (Ng et al. 2020). We estimate $RM_{iono}$ from the RMextract package[6] (Mevius 2018b), whose values are provided in Table 1.

## 3. RESULTS

### 3.1. *Burst Waterfalls and Profiles*

Table 1 reports the results of our polarimetric analysis of 44 bursts from repeating source FRB 20180916B. These bursts are summarized in Figure 1 in the form of waterfall plots (also referred to as 'dynamic spectra'), displaying each burst as a function of frequency and time. Here, data have been rebinned to highlight substructure of individual bursts. Bursts have been dedispersed by their individual structure optimizing DM, $DM_{struct}$, where structure optimization is performed on rebinned data.

Burst profiles are shown directly above waterfall plots and display the *subband* averaged signal in total (black), linear polarized (red) and circular polarized intensities (blue) as a function of time. Each burst's subband has been determined by eye and is indicated by an orange line along the left vertical axis and reported in Table 1 under the "Burst Bandwidth" column. Linear polarization, $L = \sqrt{Q^2 + U^2}$, is a positive definite quantity where noise adds coherently. As such, simply determining the linear polarized fraction using $L$ will over-estimate the true value, with the bias becoming more significant at lower S/N. We use Equation 11 of Everett & Weisberg (2001) to correct for this bias. Both the linear polarized profiles in Figure 1 and the $L/I$ values in Table 1 corresponds to the debiased measurements. Adjoining the burst profiles in Figure 1 is an additional panel showing PA behavior as a function of time, where PA values are calculated according to Equation 3.

---

[4] That said, $RM_{FDF}$ and $RM_{QU}$ largely track each other across observations.

[5] https://github.com/FRBs/pygedm

[6] https://github.com/lofar-astron/RMextract

**Table 1**: Individual burst properties of CHIME/FRB repeater FRB 20180916B.

| Burst number | Arrival Time[a] (MJD) | S/N[b] | DM$_{struct}$[c] (pc cm$^{-3}$) | n$_{down}$[d] | ⟨L/I⟩[e] | Burst Bandwidth[f] (MHz) | RM$_{FDF}$ (rad m$^{-2}$) | RM$_{QU}$ (rad m$^{-2}$) | RM$_{iono}$ (rad m$^{-2}$) |
|---|---|---|---|---|---|---|---|---|---|
| 1 | 58477.16185 | 62.2 | 348.732(10) | 32 | 0.897(20) | 400-500 | −113.59(28) | −113.65(36) | +0.06 |
| 2 | 58478.15521 | 77.9 | 348.791(23) | 8 | 0.873(16) | 485-615 | −113.43(14) | −114.14(24) | +0.14 |
| 3 | 58638.71643 | 164.6 | 348.744(32) | 8 | 0.9773(84) | 500-750 | −115.99(12) | −115.56(17) | +1.04 |
| 4 | 58639.70561 | 68.8 | 349.89(15) | 64 | 0.991(20) | 410-570 | −115.98(17) | −114.37(21) | +0.86 |
| 5 | 58639.71008 | 86.8 | 348.813(72) | 64 | 0.944(15) | 435-525 | −116.22(41) | −116.40(41) | +0.87 |
| 6 | 58786.32075 | 27.3 | 348.68(46) | 256 | 0.880(46) | 400-480 | −113.6(1.3) | −113.7(1.2) | +0.38 |
| 7 | 58835.17324 | 32.3 | 349.11(20) | 256 | 0.743(32) | 400-560 | −114.24(79) | −113.0(1.5) | +0.20 |
| 8 | 58836.17198 | 17.0 | 350.0527(75) | 256 | 1.002(83) | 400-450 | −123.3(2.3) | −121.1(3.6) | +0.26 |
| 9 | 58852.13628 | 16.7 | 349.847(83) | 256 | 0.761(64) | 630-760 | −123.0(2.6) | −123.0(5.0) | +0.23 |
| 10 | 58852.13773 | 43.9 | 349.097(61) | 64 | 0.848(27) | 400-575 | −116.18(36) | −114.85(41) | +0.22 |
| 11 | 58868.07586 | 40.5 | 348.98(12) | 128 | 0.745(26) | 525-690 | −112.05(83) | −118.3(1.1) | +0.24 |
| 12 | 58882.04681 | 16.9 | 349.27(27) | 256 | 0.933(78) | 645-770 | −114.4(2.3) | −112.4(3.7) | +0.51 |
| 13 | 58883.03995 | 32.5 | 349.37(20) | 128 | 0.941(41) | 400-470 | −117.40(64) | −116.31(70) | +0.45 |
| 14 | 58883.04405 | 22.9 | 349.725(48) | 256 | 1.031(64) | 640-735 | −112.8(2.5) | −115.6(3.8) | +0.44 |
| 15 | 58883.05372 | 28.0 | 348.840(50) | 32 | 0.636(32) | 610-735 | −115.85(91) | −114.3(1.3) | +0.41 |
| 16 | 58899.00706 | 38.5 | 348.73(25) | 128 | 0.892(33) | 400-550 | −116.89(51) | −116.90(68) | +0.47 |
| 17 | 58981.77662 | 51.7 | 349.507(84) | 64 | 0.933(26) | 650-800 | −115.22(84) | −113.69(99) | +0.92 |
| 18 | 58982.77157 | 20.8 | 349.46(57) | 256 | 0.877(60) | 400-480 | −112.1(1.2) | −114.3(1.3) | +0.93 |
| 19 | 59013.69287 | 49.2 | 348.955(31) | 32 | 0.932(27) | 400-500 | −113.19(31) | −112.49(31) | +1.01 |
| 20 | 59014.68533 | 61.5 | 348.7917(55) | 8 | 0.938(22) | 540-680 | −110.97(14) | −112.65(23) | +0.91 |
| 21 | 59111.40643 | 16.1 | 350.12(34) | 256 | 0.443(39) | 400-570 | −116.9(3.1) | −118.4(3.3) | +0.16 |
| 22 | 59225.10428 | 16.5 | 348.99(13) | 256 | 0.908(78) | 450-600 | −117.5(1.3) | −117.9(1.7) | +0.22 |
| 23 | 59241.06071 | 18.2 | 348.8435(98) | 64 | 0.911(71) | 550-730 | −116.40(92) | −117.6(1.5) | +0.43 |
| 24 | 59244.04400 | 33.8 | 348.838(71) | 256 | 0.884(37) | 420-550 | −118.87(63) | −118.02(73) | +0.49 |
| 25 | 59244.06229 | 43.9 | 348.6885(20) | 128 | 0.782(25) | 450-650 | −117.97(64) | −117.35(90) | +0.43 |
| 26 | 59245.05833 | 42.3 | 349.47(31) | 256 | 1.011(34) | 400-510 | −113.75(87) | −115.2(1.4) | +0.42 |
| 27 | 59275.96877 | 51.7 | 349.542(51) | 64 | 1.041(28) | 400-510 | −115.14(23) | −115.46(25) | +1.11 |
| 28 | 59276.96257 | 26.8 | 348.83(45) | 256 | 0.895(47) | 405-500 | −117.20(68) | −118.2(1.0) | +0.64 |
| 29 | 59277.96034 | 39.2 | 348.988(20) | 64 | 0.951(34) | 400-500 | −116.66(28) | −116.07(45) | +0.74 |
| 30 | 59306.89010 | 61.4 | 349.60(17) | 32 | 0.909(21) | 585-770 | −117.94(34) | −120.05(50) | +0.92 |
| 31 | 59355.74759 | 27.2 | 348.93(37) | 256 | 0.912(47) | 435-560 | −107.6(1.7) | −109.6(1.4) | +0.97 |
| 32 | 59357.74680 | 33.4 | 348.95(34) | 128 | 1.009(43) | 400-480 | −108.42(78) | −109.11(97) | +1.18 |
| 33 | 59390.65203 | 26.8 | 348.97(21) | 256 | 0.959(51) | 400-460 | −104.86(97) | −106.7(1.1) | +0.94 |
| 34 | 59406.60025 | 29.6 | 349.38(14) | 256 | 0.653(31) | 415-550 | −101.15(77) | −101.83(89) | +0.40 |
| 35 | 59407.59936 | 205.9 | 348.8165(93) | 1 | 0.9205(63) | 430-520 | −103.624(25) | −103.739(42) | +0.78 |
| 36 | 59440.50726 | 19.5 | 348.731(43) | 64 | 0.655(48) | 400-570 | −87.2(1.5) | −96.9(1.6) | +0.26 |
| 37 | 59440.51585 | 32.5 | 349.062(12) | 64 | 0.789(34) | 400-470 | −90.07(44) | −92.36(63) | +0.30 |
| 38 | 59486.38963 | 14.2 | 349.79(21) | 256 | 0.975(97) | 460-615 | −88.63(62) | −85.69(65) | +0.27 |
| 39 | 59519.30410 | 20.0 | 349.9(1.9) | 256 | 0.982(70) | 400-440 | −78.25(91) | −79.9(1.4) | +0.42 |
| 40 | 59519.30470 | 208.6 | 348.967(23) | 2 | 0.9702(66) | 440-540 | −83.544(42) | −78.894(86) | +0.42 |
| 41 | 59519.30909 | 72.9 | 348.862(14) | 4 | 0.827(16) | 460-565 | −81.93(19) | −79.62(28) | +0.42 |
| 42 | 59521.28871 | 60.3 | 348.94(19) | 16 | 0.932(22) | 400-450 | −80.08(26) | −80.37(36) | +0.22 |
| 43 | 59569.16266 | 28.0 | 349.05(16) | 64 | 0.764(39) | 400-430 | −67.3(1.1) | −69.5(1.3) | +0.33 |
| 44 | 59570.15498 | 100.9 | 348.901(27) | 4 | 0.891(12) | 400-535 | −69.032(88) | −69.17(12) | +0.27 |

[a] Topocentric TOAs provided here are in Modified Julian Date (MJD) format, referenced at 400 MHz with ∼ 1 second precision.

[b] The boxcar-S/N determined from downsampled data of each burst.

[c] From structure-optimization (see Section 2.2.1).

[d] The downsampling factor that determines the time resolution (i.e. $n_{down} \times 2.56 \, \mu s$) of the waterfalls plots displayed in Figure 1.

[e] The linear polarization fraction determined from integrating $L$ over the burst profile and correcting for bias.

[f] The portion of the 400-800 MHz CHIME band over which emission is observed, corresponding to vertical orange lines in burst waterfalls.

In general, the bursts studied here have high linear polarization fractions ($\gtrsim 90\%$), consistent with previous polarized observations (CHIME/FRB Collaboration et al. 2019; Chawla et al. 2020; Nimmo et al. 2021a) of this source. This suggests that the lower polarized fractions reported by recent observations with LOFAR (Pleunis et al. 2021b) are not specific to that observing epoch but, more likely, reflect a frequency dependence in the linear polarized signal; one that is possibly an imprint of a depolarization effect (see Section 4.2). Meanwhile, a low level circular polarization is present in the sample but is likely an artefact of cross-polarization leakage from imperfect feed alignment rather than astrophysical signal as has been observed in other bursts reported from CHIME/FRB (e.g., Fonseca et al. 2020; Mckinven et al. 2021). This interpretation is supported by the greater circular polarization in brighter bursts where cross-polarization should be more evident. The observed behavior in PA matches previous observations, namely a relatively flat PA curve across the burst duration with subtle short timescale variability recently noted for this source by Nimmo et al. (2021a).

### 3.2. *Burst Properties vs. Time*

Figure 2 displays the individual burst properties of the 44-burst sample spanning 2018 December 25 to 2021 December 22. Properties displayed are the RM (panel A), $DM_{struct}$ (panel B), fractional linear polarization ($L/I$; panel C) and the bandwidth (panel D) of each burst. Panel E displays each burst's phase location in the 16.33 periodic cycle of this source, represented as vertical blue, dotted lines on the panel. Significant RM evolution is seen from this source, particularly between 2021 April and 2021 December where the RM has changed by $\sim 50 \, \mathrm{rad \, m^{-2}}$. This is does not appear to be matched by any correlated evolution in any of the other properties. RMs displayed here refer to $RM_{QU}$ (blue markers) after correcting for ionospheric ($RM_{iono}$) contributions and are compared to previously published RMs obtained from LOFAR observations spanning $110 - 188$ MHz ( green triangles; Pleunis et al. 2021b) and two separate observations with the GBT; one covering $300 - 400$ MHz (magenta squares; Chawla et al. 2020) and another spanning $680 - 920$ MHz (cyan hexagons; Sand et al. 2021)[7].

The observed RM variations greatly exceed the errors introduced by ionospheric corrections or residual systematics, which combined should not exceed $\sim 1 \, \mathrm{rad \, m^{-2}}$. The characteristic $\lambda^2$ scaling of Faraday rotation cannot be easily replicated by instrumental effects, protecting the robustness of RM measurements [8]. The source of the RM variability reported here for FRB 20180916B is therefore astrophysical. This is strongly supported by the analysis of high S/N events for which discrepant RM measurements are most obvious. Figure A.1 displays FDFs of our 44-burst sample. Each FDF has been normalized by its peak value to ease comparison across a sample that explores a wide range of burst S/N. $RM_{QU}$ measurements are indicated as vertical green lines and largely correlate with the peak locations of the FDFs.

The RM evolution reported in Figure 2 appears to have two distinct regimes. From 2018 December to April 2021, RM variations are small but significant, with weak evidence for clustering on weeks-months timescales. This quasi-random behavior is followed by an interval over 2021 April−December when $|RM|$ exhibits a sustained decrease of $\sim 50 \, \mathrm{rad \, m^{-2}}$. The most recent RM measurements now suggest a possible change in sign of $RM_{excess}$ similar to the recent observations of FRB 20190520B (Anna-Thomas et al. 2022). This result may suggest a reversal in the orientation of $B_\parallel$. However, the Galactic RM contribution is highly uncertain in this direction and most recent RM measurements are still within the error budget of the $RM_{MW}$ estimate.

The 2021 April−December epoch is remarkably well-described by a linear trend in RM versus time. Performing a linear regression yields a gradient in the RM of $d\mathrm{RM}/\mathrm{dt} = 0.197 \pm 0.006 \, \mathrm{rad \, m^{-2} \, day^{-1}}$ with a reduced chi-square statistic of $\chi_\nu^2 = 2.5$. Such quasi-linear behavior is reminiscent of similar trends seen in FRB 20121102A (Hilmarsson et al. 2021b) and the Galactic center magnetar, PSR J1745−2900 (Desvignes et al. 2018). However, unlike FRB 20121102A that exhibits a secularly increasing DM of $\gtrsim 1 \, \mathrm{pc \, cm^{-3} \, year^{-1}}$, no correlated evolution in $DM_{struct}$ is observed from FRB 20180916B. This either implies that variations in RM are dominated by fluctuations in $B_\parallel$ or that our $DM_{struct}$ measurements are not sufficiently sensitive to capture the modest but coherent DM evolution required to produce the observed RM evolution. Regardless of the fidelity of $DM_{struct}$ measurements, changes in the local environment are required to explain the magnitude of the RM variations and short timescales over which they occur. The linearly increasing RM evolution of this source suggests an imminent and unambiguous change in sign of $RM_{excess}$ if the trend continues over the next few months. Such behavior would be reminiscent of the RM sign changes recently observed from FRB 20190520B and would imply that the observed RM evolution is dominated by changes in $B_\parallel$ of local environment rather than changes in the free electron column depth.

---

[7] We omit the RMs reported from Nimmo et al. (2021a) due to the large uncertainties reported therein.

[8] This is particularly true at larger $|RM|$ and fractional bandwidths, where the $\lambda^2$ scaling of Faraday rotation is less likely to be confused with residual systematics.

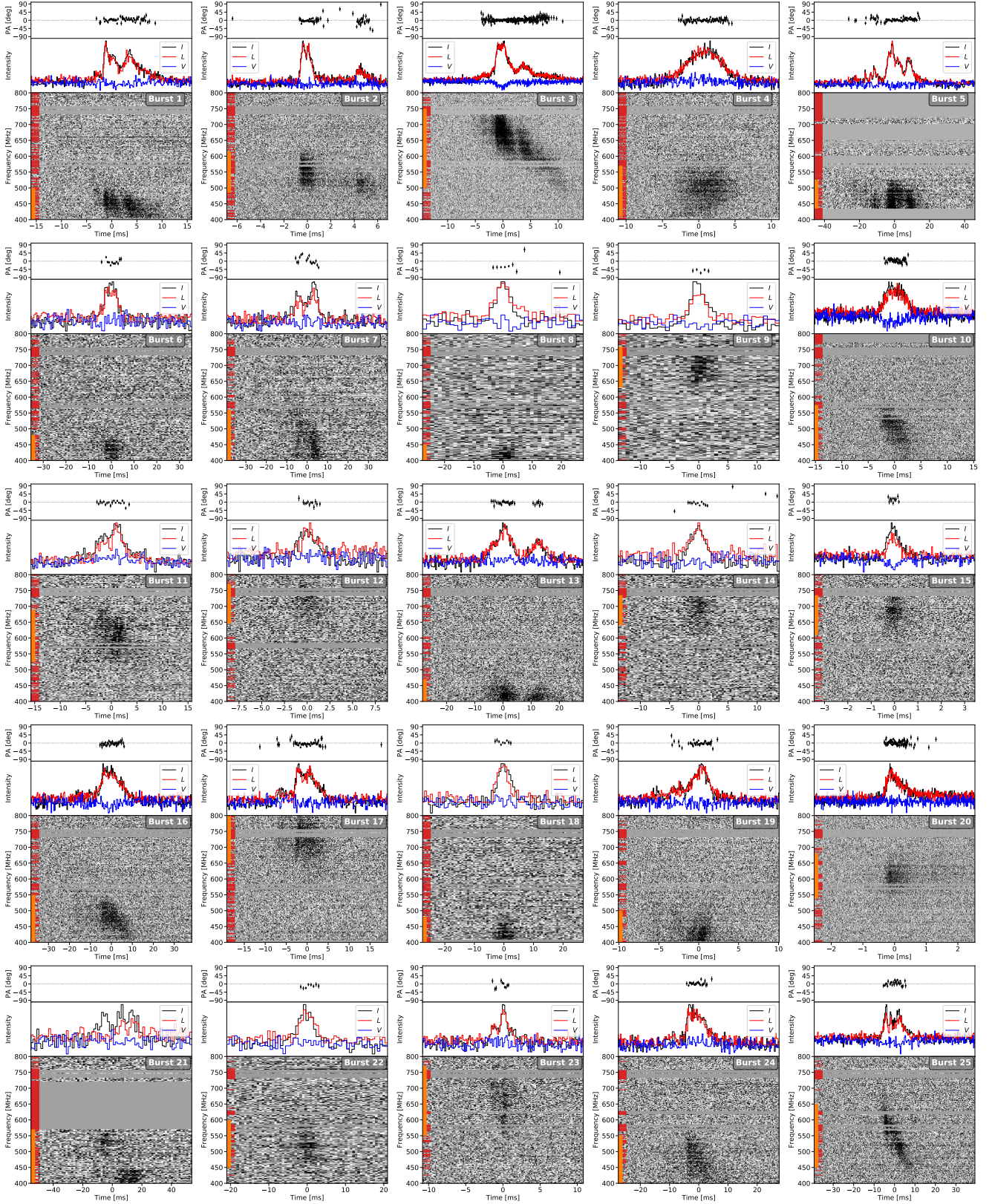

**Figure 1**: Waterfall plots in Stokes I of bursts from repeating source FRB 20180916B in chronological order and dedispersed to their structure optimized DMs (DM$_{\rm struct}$; listed in Table 1). Data have been rebinned to a fixed frequency resolution of $\nu_{\rm res} = 1.5625$ MHz. Time has been rebinned according to $n_{down}$ of each burst, such that $t_{\rm res} = n_{down} \times 2.56~\mu$s. Panels above the spectra display burst profiles of total (black), linear polarized (red), and circularly polarized (blue) intensity (peak normalized), as well as the polarization angle (PA). Burst profiles are obtained by adding signal over the spectral limits of the burst, indicated by orange lines along the frequency axis. Masked frequency channels are indicated by red lines along the vertical axis. Each panel is labeled with the corresponding burst number from Table 1.

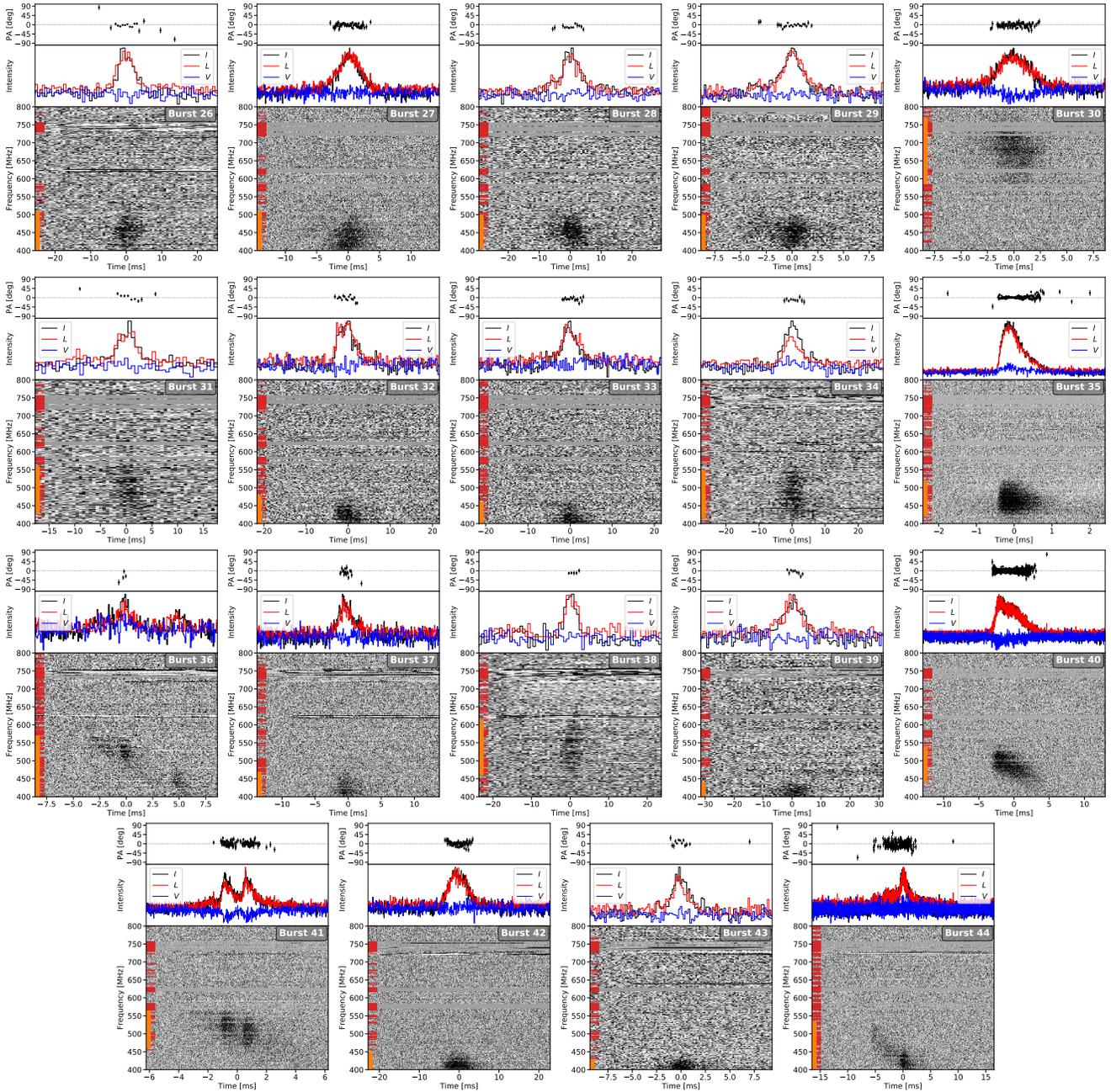

**Figure 1**: (Continued) Waterfall plots of bursts from repeating source FRB 20180916B.

Assuming that the observed RM evolution is entirely an imprint of a changing $B_\parallel$, we estimate a lower limit on the average rate of change of $B_\parallel$,

$$\langle dB_\parallel/dt \rangle = 1.23 \frac{d\mathrm{RM}/dt}{\mathrm{DM}_{\Delta\mathrm{RM}}}(1+z) \gtrsim 0.004 \ \mu\mathrm{G\,day^{-1}}. \quad (4)$$

Here, $d\mathrm{RM}/dt$ corresponds to the slope of linear-fit over the secular regime (orange line in Fig. 2) while $\mathrm{DM}_{\Delta\mathrm{RM}}$ is the unknown DM contribution of the Faraday-active medium.

We arrive at a conservative lower limit for $\langle dB_\parallel/dt \rangle$ by assuming $\mathrm{DM}_{\Delta\mathrm{RM}} = 70 \ \mathrm{pc\,cm^{-3}}$, the upper limit of the host galaxy DM contribution for this source (Marcote et al. 2020). A $(1 + z)$ term is included to account for the differential dilution of cosmological expansion on DM and RM, however, this does not significantly affect the estimate given the low redshift ($z = 0.0337$; Marcote et al. 2020) of this source. We emphasize that the $\mathrm{DM}_{\Delta\mathrm{RM}}$ value assumed here very likely overestimates the DM contribution of the Faraday-active medium and thus it remains plausible that the true $dB_\parallel/dt$ significantly exceeds the lower limit. For instance, $dB_\parallel/dt \sim 0.4~\mu$G is required if we assume that only the variable component of the DM inferred from our $DM_{struct}$ measurements (i.e., $\Delta DM = 0.8~\mathrm{pc\,cm^{-3}}$), contributes to the observed RM evolution of this source. Interpretations of our $dB_\parallel/dt$ measurement as well as possible methods for further constraining it are provided in Section 4.1.

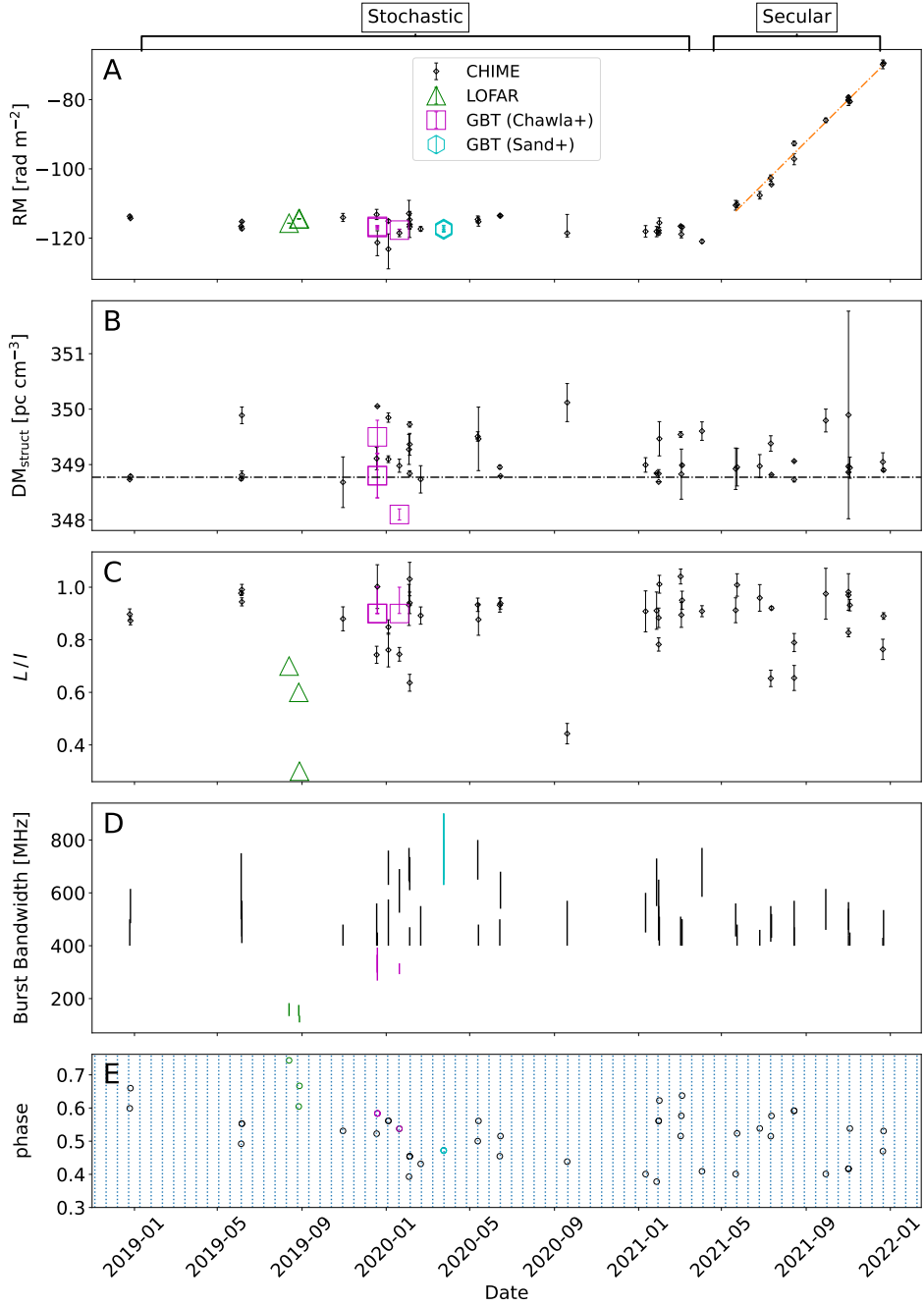

**Figure 2**: Burst properties as a function of time for baseband data recorded from FRB 20180916B, displaying ionospheric corrected RMs (panel A), $DM_{struct}$ (panel B), linear polarized fraction ($L/I$; panel C) and burst bandwidth (panel D) and phase location within the 16.33 day activity cycle (panel E). Times are in Coordinated Universal Time (UTC) format and phases are referenced relative to $\phi_0 = 58369.40$ MJD such that the vertical blue-dotted lines in panel E correspond to the mean of the folded phases of the 44-burst sample. The RMs of Panel A (black points) are determined from QU-fitting ($RM_{QU}$) applied to the 44-burst sample and are displayed alongside previously published RMs obtained from LOFAR (Pleunis et al. 2021b) and several GBT observations (Chawla et al. 2020; Sand et al. 2021). An apparent secular decrease in $|RM|$ from $RM \sim -120\ \mathrm{rad\,m^{-2}}$ to $\sim -70\ \mathrm{rad\,m^{-2}}$ over 2021 April - 2021 December is highlighted by an orange line with best-fit slope of $0.197 \pm 0.006\ \mathrm{rad\,m^{-2}\,day^{-1}}$. The trend is consistent with a decreasing $|RM_{excess}|$ contribution but the precise value remains highly uncertain given the substantial error in the $RM_{MW}$ contribution in the direction of this source (Section 2.3). The horizontal, dashed-dotted line in Panel B corresponds to the structure optimized value $DM = 348.772\ \mathrm{pc\,cm^{-3}}$ determined by Nimmo et al. (2021a) using high time resolution observations from the European VLBI Network (EVN). For Panel C, data points where $L/I > 1$ are nonphysical but are consistent with expectations for a 100% linearly polarized source and the reported measurement uncertainties.

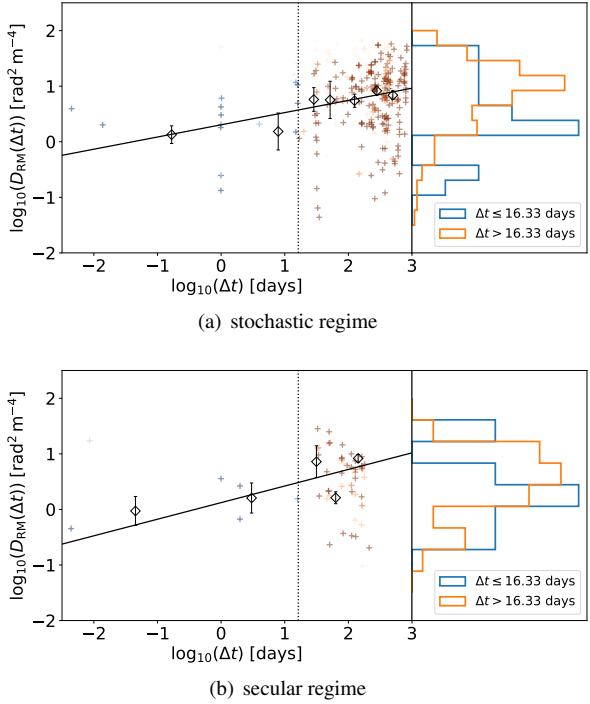

(a) stochastic regime

(b) secular regime

**Figure 3**: The RM structure functions (corrected for measurement errors) for the two evolutionary regimes of FRB 20180916B; an epoch in which the source displays apparently stochastic RM evolution (2018 December - 2021 March; Panel a) and another exhibiting secular RM evolution (2021 April - 2021 December; Panel b). The vertical dotted line indicates the 16.33 day activity cycle of this source. $D_{RM}(\Delta t)$ measurements are indicated by cross markers and divided into two samples based on their $\Delta t$ value: $\Delta t \leq 16.33$ days (blue) and $\Delta t > 16.33$ days (orange). For both samples, a saturating color scale emphasizes data points with more significant values relative to the noise contributed by RM measurement errors. Black markers indicate rebinned data and correspond to the mean values with uncertainties estimated as the standard error on the mean. Black lines correspond to power-law fits, $D_{RM}(\Delta t) \sim \Delta t^l$, such that, $l \sim 0.22$ (stochastic) and $l \sim 0.30$ (secular). The adjoining histograms display the $D_{RM}(\Delta t)$ distribution of the two samples. $D_{RM}(\Delta t)$ values in panel (b) have been calculated after de-trending the secular RM evolution using the linear-fit displayed in Figure 2.

### 3.3. *Structure Function Analysis*

The RM variability of FRB 20180916B is dominated by the secular evolution displayed between 2021 April and 2021 December. However, significant RM variations are observed even at times before this epoch, when RM evolution does not follow any predictable trend. We employ structure function analysis to more systematically study this RM variability, a useful tool for probing correlated behavior in situations where the underlying distribution is sparsely and/or irregularly sampled (Schulz-Dubois & Rehberg 1981). Modifying the notation of Minter & Spangler (1996) for our time series data, the structure function of RM variations can be expressed as,

$$D_{RM}(\Delta t) = \frac{1}{N} \sum_i [RM(t) - RM(t + \Delta t)]_i^2 \quad (5)$$

Here, the sum is over all pairs of observations with a time separation of $\Delta t$ and $N$ is the number of pairs included in the sum. $D_{RM}(\Delta t)$, therefore, represents the variance in $\Delta RM$ at different time separations. For a perfectly random and stationary time series, the structure function, $D_{RM}(\Delta t)$, would be uniform as a function of $\Delta t$. Any excursion from uniformity is therefore evidence for correlated behavior as a function of time with the specifics of the distribution providing further insight into the dynamics/geometry of the system.

We apply Equation 5 to our 44-burst sample, splitting our sample into two non-overlapping epochs: 1) 2018 December to 2021 April ($n = 30$; stochastic regime) and 2) 2021 April to 2021 December ($n = 14$; secular regime). These results are displayed[9] in Figure 3. $\log_{10}(D_{RM}(\Delta t))$ measurements are displayed as cross markers and have been corrected for the noise bias introduced by non-uniformity in measurement errors. This is done by subtracting $\sigma^2_{RM(t_i)} + \sigma^2_{RM(t_j)}$, where $\sigma_{RM(t_{i,j})}$ are the measurement errors for a pair with some separation $\Delta t = t_i - t_j$. The vertical dotted line indicates the 16.33 day period of this source and measurements are colored blue ($\Delta t \leq 16.33$ days) or orange ($\Delta t > 16.33$ days) in relation to this benchmark timescale. The (normalized) distributions of $\log_{10}(D_{RM})$ of these two samples are displayed in adjoining panels. $D_{RM}(\Delta t)$ values for the secular regime (panel b) have been calculated after using a linear fit (orange line; Figure 2) to de-trend the secular RM evolution. For both samples, a saturating color scale is used to emphasize more significant measurements relative to their errors.

For both the stochastic and secular regimes, we find a possible weak dependence of $D_{RM}$ on $\Delta t$ in which RM variability increases with larger $\Delta t$. We estimate the significance of this trend by performing a two-sample Kolmogorov-Smirnov (KS) and Anderson-Darling (AD) test on $\log_{10}(D_{RM})$ for $\Delta t \leq 16.33$ and $\Delta t > 16.33$. For the stochastic regime, we find a marginally significant difference between the two distributions with a p-value falling in the range of 0.04 (AD-test) and 0.06 (KS-test)[10]. Apart from this weak trend, we do not see any significant evidence of a systematic change in $\Delta t$ at 16.33 days or any other timescale. This suggests that the trend, to the limits of our RM measurement precision, is not related to the mechanism generating the periodic activity but simply a reflection of inhomogeneities in the source's local magneto-ionic environment. Alternatively, rather than reflecting astrophysical origins, the weak trend between $D_{RM}$ and $\Delta t$ may be a product of limited sampling, incorrect measurement errors, and artefacts from imperfect ionospheric correction could produce the observed behavior.

---

[9] The non-uniform sampling of $\Delta t$ is an imprint of the periodic activity cycle of the source, which causes $\langle D_{RM}(\Delta t)\rangle$ measurements to cluster at increments of 16.33 days.

We applied the same procedure to $DM_{struct}$ and $\langle L/I \rangle$ measurements and find no evidence for a dependence of either of these quantities on $\Delta t$ for both the stochastic and secular regimes. In the case of DM, the low S/N and narrow bandwidth of many bursts of our sample makes it highly likely that our observations are insensitive to small intrinsic DM variability of this source. Furthermore, our $DM_{struct}$ uncertainties are very likely underestimated and skewed by confusion with burst morphology (see Section 4.5). This is supported by the clustering of our higher S/N bursts near the reference value $DM = 348.772 \, pc \, cm^{-3}$ that was independently determined with high time-resolution data (Nimmo et al. 2021a). For these reasons, we do not place too much emphasis on our $D_{DM}(\Delta t)$ measurements.

### 3.4. *Spectral Evolution between the Stochastic and Secular RM Regimes*

The interval of time between 2021 April and 2021 December is accompanied by a subtle change in the emitting band of the source, in which emission appears to have drifted to lower frequencies. This result is summarized in Figure 4 which shows the distributions of the emitting band center of bursts that occur during the stochastic and secular regimes of RM evolution. A two sample AD-test produces a marginally significant result (p-value = 0.02). This significance is likely underestimated since the secular regime contains a higher fraction of bursts that are band-limited (i.e. emission persists at frequencies below 400-800 MHz bandpass of CHIME). This interval is not associated with a systematic change in activity cycle phase location, indicating this possible evolution in the burst spectra is not related to the known dependence on burst activity of this source with frequency (Pastor-Marazuela et al. 2020; Pleunis et al. 2021b). Future multiband observations of this source will help determine the significance of this result. If this relation persists, then this would suggest a common mechanism for generating the secular RM and burst spectral evolution.

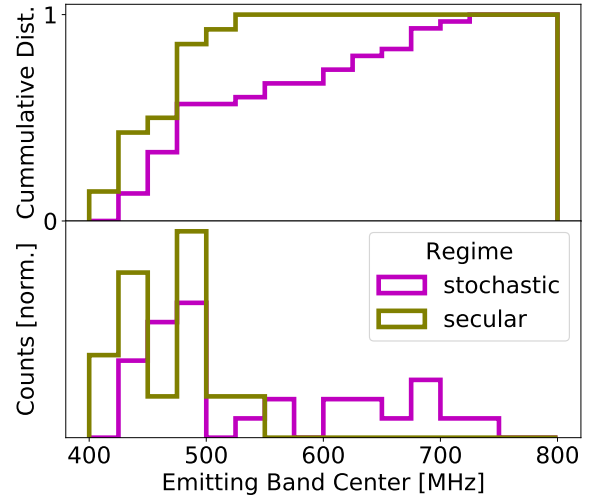

**Figure 4**: The (normalized) distribution of the emitting band center of bursts detected from FRB 20180916B during the stochastic ($n = 30$; magenta) and secular ($n = 14$; gold) regimes of RM evolution. The two distributions are marginally discrepant (see text for details) with bursts that occur during the secular regime preferentially centered on the bottom half ($400 - 600$ MHz) of the CHIME band.

### 3.5. *RM vs. Activity Cycle Phase*

Recent observations with LOFAR indicated a possible dependence of RM with phase location in the 16.33 activity cycle of FRB 20180916B (Pleunis et al. 2021b). While this remains a possibility, the observations shown here demonstrate that substantial RM variability exists across activity cycles, thus posing a challenge for identifying an imprint of phase on the observed RM. In particular, inter-cycle RM variations, whether they be stochastic or secular, will decorrelate a hypothetical dependence of RM with phase location. We explored detrending this slow timescale variability but found results to be highly sensitive to provided inputs of the detrending analysis[11].

The secular regime from 2021 April to 2021 December offers the best opportunity for probing a subtle relation between RM and phase location by allowing us to detrend the broadly linear RM evolution displayed over this interval. Figure A.3 demonstrates the results of this analysis where residuals, $RM - f_{RM}(t)$, are relative to the linear fit displayed in Figure 2. Excursions in the residual RM are significant, however, there is no clear evidence for a relation of

---

RM with phase location. Furthermore, we calculate a root-mean-square (RMS) of $\sim 1.5\ \mathrm{rad\,m^{-2}}$ for these residuals, which is significantly smaller than the equivalent RMS of $\sim 2.5\ \mathrm{rad\,m^{-2}}$ obtained over the stochastic regime. This result, combined with the analysis of Section 3.3, suggests that the mechanism generating the periodic activity of this source remains broadly insensitive to the magneto-ionic fluctuations that produce the RM variations. However, given limitations of our RM measurement precision, it remains possible that subtle ($\sim 1\ \mathrm{rad\,m^{-2}}$) RM dependence may be tied to the activity cycle of the source. This analysis should be revisited in the future with a much larger sample of bursts, potentially enabling a subtle relation of RM with phase to be parsed from stochastic contributions.

## 4. DISCUSSION

We summarize the main results of our analysis below.

1. Over three years, CHIME/FRB recorded baseband data for 44 bursts from periodically active source FRB 20180916B – a factor of four increase in the published sample of bursts with polarization information for this source.

2. Bursts from FRB 20180916B regularly display linear polarized fractions between $80-100\%$ with relatively flat PA curves exhibiting subtle variations on short-timescales ($\ll 1$ ms). Some bursts display evidence of circular polarization that is likely to be substantially instrumental.

3. FRB 20180916B exhibits significant time dependent variations in RM that include an epoch where RM variability is (quasi-) stochastic (2018 December−2021 April) and another in which it is secular (2021 April−2021 December). During the secular regime, the RM was observed to increase by approximately $50\ \mathrm{rad\,m^{-2}}$ to its most recent value of RM $\sim -70\ \mathrm{rad\,m^{-2}}$, which is now within the errors of the $\mathrm{RM_{MW}}$ estimate of this source.

4. The secular regime is well described by a linear RM trend with stochastic DM variations of $\leq 0.8\ \mathrm{pc\,cm^{-3}}$, indicating a changing local magnetic field such that, $\langle dB/dt \rangle \gtrsim 0.004\ \mu\mathrm{G\,day^{-1}}$. Burst emission during this interval is over-represented at lower frequencies, indicating a possible relation between the emitting band and RM evolution.

5. Applying structure function analysis, we find possible evidence for a trend of increasing RM variability for greater time separation between bursts. RM variations of bursts that occur within the same activity cycle ($\Delta t \leq 16.33$ days) are, on average, smaller than those occurring at different cycles ($\Delta t > 16.33$ days).

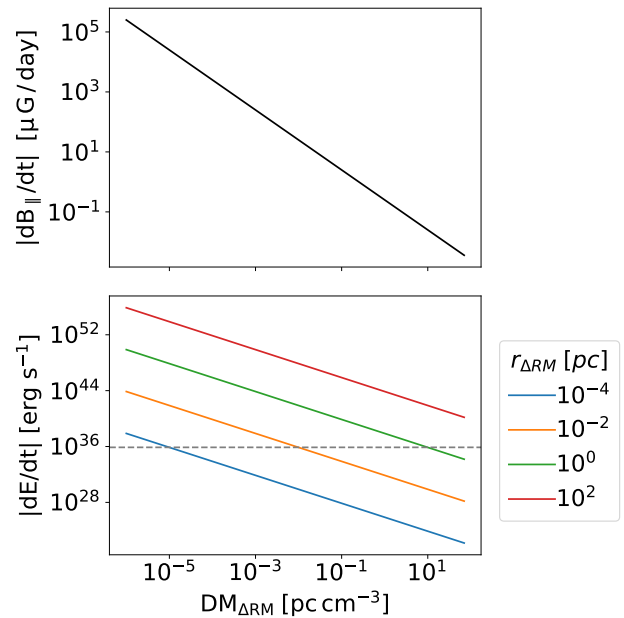

**Figure 5**: Top panel: The absolute rate of change in the LoS component of the magnetic field, $|dB_\parallel/dt|$, as a function of the DM contribution of the Faraday-active medium, $\mathrm{DM_{\Delta RM}}$, of FRB 20180916B. The maximum $\mathrm{DM_{\Delta RM}}$ value is constrained by the upper limit on the DM contribution of the host galaxy ($\mathrm{DM_{host}} \lesssim 70\ \mathrm{pc\,cm^{-3}}$; Marcote et al. 2020). Bottom panel: The associated absolute rate of energy loss, $|dE/dt|$, for different spatial scales of the Faraday-active medium ($\mathrm{r_{\Delta RM}}$). The grey horizontal line indicates the current upper limit on the luminosity of a persistent radio counterpart of FRB 20180916B ($\mu L_\nu \leq 7.6 \times 10^{35}\ \mathrm{erg\,s^{-1}}$; Marcote et al. 2020).

### 4.1. *Interpretation of RM Evolution*

A significant fraction of repeating FRB sources appears to occupy complex magneto-ionic environments as evidenced by the significant RM evolution displayed from many of these sources (e.g., Hilmarsson et al. 2021b; Xu et al. 2021; Anna-Thomas et al. 2022; Dai et al. 2022). FRB 20180916B, studied here, appears to be no exception despite its relatively modest $|\mathrm{RM_{excess}}|$. Such a result indicates that significant fluctuations in the local magneto-ionic environment can exist even for sources that do not appear to reside in highly magnetized environments. In general, the characteristic RM variability seen from repeating FRB sources may reflect any combination of changes in the free electron density ($n_e$), magnetic field ($\vec{B}$) and the distance and/or geometry through the Faraday-active medium. Without additional information a variety of competing scenarios may be invoked to explain RM variability of a source.

In the case of FRB 20180916B, the absence of any significant correlated behavior between DM and RM strongly suggests that the observed RM evolution of this source is produced by changes in the Faraday-active medium's magnetic field. In Section 3.2 we estimated a lower limit on the rate of change in the LoS component of the magnetic field ($dB_\parallel/dt$) by assuming a DM contribution of the Faraday-active medium ($\mathrm{DM_{\Delta RM}}$) that equaled the upper limit of the DM contribution of the host galaxy ($\mathrm{DM_{host}} \lesssim 70$ pc cm$^{-3}$; Marcote et al. 2020). If we relax this condition, such that $\mathrm{DM_{\Delta RM}} \ll 70$ pc cm$^{-3}$, we obtain significantly larger $dB_\parallel/dt$ values. Figure 5 shows the dependence between $dB_\parallel/dt$ and $\mathrm{DM_{\Delta RM}}$ and demonstrates that $dB_\parallel/dt$ values spanning several orders of magnitude are plausible depending on constraints given to $\mathrm{DM_{\Delta RM}}$.

If we assume that the $dB_\parallel/dt$ values reflect a change in the strength of the magnetic field, we can provide a rough estimate of the associated energy change,

$$\Delta E = \frac{B_{t0}^2 - B_{t1}^2}{8\pi} \times \text{Volume}. \quad (6)$$

Here, we assume that the Faraday-active medium of a given volume is uniformly permeated by a magnetic field of changing strength. $B_{t0}$ and $B_{t1}$ represents the magnetic field strength at the beginning and end of an interval of time. In our case, we use the 9 month interval of the secular RM regime and estimate $B_{t0}$ and $B_{t1}$ from our $dB_\parallel/dt$ estimates displayed in Figure 5. In particular, $B_{t0}$ and $B_{t1}$ are calculated for each $\mathrm{DM_{\Delta RM}}$ such that $B_{t0}$ is estimated using the $|\mathrm{RM}|$ at the beginning of the secular regime which is then extrapolated using the corresponding $dB_\parallel/dt$ estimate to obtain $B_{t1}$. Importantly, this calculation assumes that $dB_\parallel/dt$ estimates reflect a change in strength of the magnetic field and not orientation and, as such, $dB_\parallel/dt \sim dB/dt$. Normalizing $\Delta E$ by the associated secular regime timespan provides us with an average rate of energy change, $|dE/dt|$, of the Faraday-active medium. The top panel of Figure 5 displays $|dE/dt|$ as a function of $\mathrm{DM_{\Delta RM}}$ for different spatial scales ($\mathrm{r_{\Delta RM}}$) of the Faraday-active medium, where we have assumed a uniform spherical medium. $|dE/dt|$ values increase for smaller $\mathrm{DM_{\Delta RM}}$ values but larger physical scales, $\mathrm{r_{\Delta RM}}$.

Given current uncertainties in the $\mathrm{RM_{MW}}$ contribution of this source, it remains unknown whether $|\mathrm{RM_{excess}}|$ is increasing or decreasing over the secular regime. The $|dE/dt|$ trendlines of Figure 5 are represented so as to be agnostic to these two possibilities. If, however, we assume a dissipating magnetic field (i.e., $dB/dt < 0$) then the associated energy loss may be observable as a radio counterpart. In such a scenario, we can place upper limits on the size of the Faraday-active medium by requiring $|dE/dt|$ estimates to be below the current radio counterpart upper limit for this source. Using the $\nu L_\nu < 7.6 \times 10^{35}$ erg s$^{-1}$ as an upper limit

(Marcote et al. 2020) requires $\mathrm{r_{\Delta RM}} \lesssim 4$ pc. This upper limit is greater than the $\sim 1$ pc upper limit on the compact radio counterpart of FRB 201201102A (Plavin et al. 2022) and broadly consistent with spatial scales of only very young ($< 1000$ years) supernova remnants. The upper limit may be improved in the future if $\mathrm{DM_{\Delta RM}}$ is further constrained. For example, DM variability of repeating FRBs is regularly $\mathcal{O}(1\,\mathrm{pc\,cm^{-3}})$. If we assume $\mathrm{DM_{\Delta RM}} = 1$ pc cm$^{-3}$ for this source then this would require $\mathrm{r_{\Delta RM}} < 0.2$ pc. Such spatial scales are significantly below $30 - 60$ pc resolution achievable for this source with current instruments (e.g., Tendulkar et al. 2021) and may suggest that the Faraday-active region is located in the immediate environment of the emission region as might be expected near the termination shock of a synchrotron blast wave model ($\sim 10^{17}$ cm; Metzger et al. 2019) or at even smaller scales in models invoking tight orbits with high mass stellar companions (e.g., Lyutikov et al. 2020; Sridhar et al. 2021).

Alternatively, the $dB_\parallel/dt$ required to produce RM evolution of FRB 20180916B may reflect inhomogeneities in a magnetized foreground structure that transits across the LoS. In this framework, $dB_\parallel/dt$ is an imprint of a spatial dependence in the Faraday-active medium rather than intrinsic changes in the magnetic field strength. Recently, Wang et al. (2022) have demonstrated that the general RM evolution of FRB 20201124A and FRB 20190520B can be reproduced by a binary system in which an FRB-emitting magnetar orbits a Be star companion with a "decretion" disk. Within this model, RM variations are an imprint of the magnetized decretion disk which is differentially probed during periastron passage of the background magnetar. Importantly, a decretion disc with an azimuthally directed B-field should result in a characteristic sign change in the RM during periastron passage, a product of the opposite directions of the decretion disc's B-field relative to the source during the ingress and egress of the magnetar's orbit.

Such a model may explain the possible sign change in $\mathrm{RM_{excess}}$ observed here from FRB 20180916B. We explore this possibility by fitting the secular RM evolution of FRB 20180916B with a simple model in which the observed RM changes are entirely due to a rotation of the magnetic field relative to the LoS. We find that the best-fit model is able to reproduce the quasi-linear RM evolution displayed over the secular regime and requires the magnetic field to rotate $\sim 100°$ over this interval, the equivalent of $\sim 0.4°$ day$^{-1}$. Whether or not a reversal in the orientation of the B-field has occurred remains ambiguous given the large uncertainty in the $\mathrm{RM_{MW}}$ contribution for this source (Section 2.3). Regardless, if the RM evolution of FRB 20180916B is indeed produced by its orbit around Be-star companion, the period of the orbit must exceed the three-year interval of our observations. This significantly exceeds the periodic activity cycle of this source that initially motivated consideration of orbital models (see Section 4.4). Therefore, if this source's RM evolution is an imprint of an orbit, an alternative mechanism may be required to explain this source's periodic bursting behavior. An orbital period in excess of three years would be substantially larger than the 80 day and 600 day orbital periods estimated for FRB 20201124A and FRB 20190520B (Wang et al. 2022) and would be on the high end of orbital periods observed from the Galactic sample of Be-type high-mass X-ray binaries (Karino 2021). Continued polarimetric monitoring of FRB 20180916B over the next year will be important for definitively detecting the presence of a field reversal and will be crucial input for models seeking to explain the RM evolution of this source.

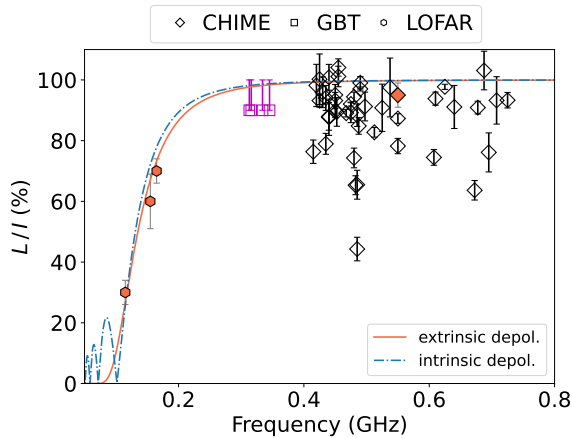

**Figure 6**: Degree of linear polarization as a function of frequency for multiband observations of FRB 20180916B. Previous measurements (orange data points) have been updated to include GBT measurements (magenta squares; Chawla et al. 2020) and those reported here with CHIME (black diamonds). The best-fit external depolarization model of Feng et al. (2022) is indicated by an orange line ($\sigma_{RM} = 0.12 \ \mathrm{rad \, m^{-2}}$) and is contrasted with a competing internal depolarization model (blue line; see text for details). Figure updated and adapted from Feng et al. (2022).

### 4.2. Depolarization Effects

Recently, Feng et al. (2022) studied multiband observations from a handful of repeaters and found $L/I$ measurements to be systematically lower at lower frequencies. This result was shown to be well described by a depolarization model that invokes an RM scatter term, $\sigma_{RM}$, to quantify the amount of stochastic Faraday rotation produced from multi-path propagation through a non-uniform magneto-ionic medium. This model was applied to observations of FRB 20180916B, which at the time included a single burst detection with CHIME (400-800 MHz; CHIME/FRB Collab-oration et al. 2019) and three others from LOFAR (110-188 MHz; Pleunis et al. 2021b). Figure 6 revisits this analysis, updating $L/I$ measurements to include the sample studied here, along with those determined with GBT (300-400 MHz; Chawla et al. 2020). In general, these additional $L/I$ measurements are significantly larger than those previously observed by LOFAR. While this result is consistent with expectations for a depolarization model, the high degree of scatter observed in our $L/I$ measurements indicates that depolarization is unlikely to be the only factor influencing $L/I$ measurements.

One possibility is that the observed scatter in $L/I$ reflects variations that are intrinsic to the source. Indeed, single pulse studies of several millisecond pulsars have demonstrated a tendency for the brightest single pulses to be the most significantly polarized (Osłowski et al. 2014; Liu et al. 2016; Feng et al. 2021). This feature has also been seen in observations of PSR J0540−6919, the second-most energetic radio pulsar known and one observed to emit giant pulses (Geyer et al. 2021). This relation between polarization and energetics can be significant, with average $L/I$ measurements of the brightest pulses reliably $\gtrsim 20\%$ larger than the mean integrated value. We are not aware of any theoretical models explaining this dependence between polarization and energetics, however, this feature may explain the significant scatter in our $L/I$ measurements from FRB 20180916B. In the absence of a precise characterization of the day-to-day variations in system sensitivity, our characterization of burst flux/fluences remain uncertain and we defer such analysis to future work.

Another possibility for explaining the scatter in $L/I$ measurements of FRB 20180916B is that the intrinsic burst emission is consistently ∼ 100% linearly polarized but that the observed scatter in $L/I$ is a consequence of a changing $\sigma_{RM}$. If this were the case, $\sigma_{RM}$ would need vary between values as low as $0.12 \ \mathrm{rad \, m^{-2}}$ (orange line in Figure 6; Feng et al. 2022) to as high as ∼ $16 \ \mathrm{rad \, m^{-2}}$. This is an unlikely scenario given that the RM of this source changes no more than ∼ 50% over the duration of our observations. That said, this scenario can be tested in the future by determining the level of correlation between $L/I$ (or $\sigma_{RM}$) and scattering timescale.

A final possibility is that an alternate/additional depolarization effect may be producing the observed scatter in $L/I$ for this source. The depolarization model used by Feng et al. (2022), referred to elsewhere as "stochastic Faraday rotation" (Melrose & Macquart 1998), is an example of an external depolarization process wherein signal is depolarized from the random RM-scatter, $\sigma_{RM}$, introduced by the foreground to the emitting region. In this scenario, depolarization monotonically increases at longer wavelengths such that the fractional reduction in the linear polarization amplitude, $f_{ext}$, can be expressed as,

$$f_{\text{ext}} = 1 - \exp(-2\lambda^4 \sigma_{\text{RM}}^2). \tag{7}$$

Alternatively, under certain conditions, depolarization can also occur within the emitting region. Such a scenario is referred to as internal depolarization and commonly occurs in circumstances where the emitting region is spatially extended, such as the diffuse, synchrotron emission from the Milky Way's interstellar medium (ISM) (e.g., Gardner & Whiteoak 1966; Burn 1966; Sokoloff et al. 1998; Brentjens & de Bruyn 2005). In this scenario, the emitting region is co-located with the medium producing Faraday rotation. This produces differential Faraday rotation as a function of path length through the emitting volume that, if significant enough, partially depolarizes the emission. Crucially, this scenario is different from external depolarization described earlier in that no relation to temporal scattering is required to produce depolarization. In the simplest possible scenario, the emitting region can be described as a uniform slab permeated by a regular magnetic field. In this case, the fractional reduction in the linear polarization amplitude, $f_{\text{int}}$, can be expressed as,

$$f_{\text{int}} = 1 - \frac{\sin(2\phi_{\text{emit}}\lambda^2)}{2\phi_{\text{emit}}\lambda^2}, \tag{8}$$

where $\phi_{\text{emit}}$ corresponds to the Faraday width of the emitting region.

In the case of FRBs, microsecond structure observed in the burst profiles of this and other sources limits the size of the emitting region, and thus imposes strong constraints the amount of differential Faraday rotation that can occur. While internal depolarization may seem disfavored by the short time scales of FRB emission, recent observations of FRB 20121102A found linear polarized fractions that agreed remarkably well with the predicted depolarization if the Faraday width of the burst environment was $\sim 150 \ \text{rad m}^{-2}$ (Plavin et al. 2022). We apply the internal depolarization model of Equation 8 to $L/I$ measurements from FRB 20180916B and retrieve a best-fit model that gives a Faraday width of $0.18 \ \text{rad m}^{-2}$. This model, displayed in Figure 6 (blue line), is able to replicate the trend of decreasing $L/I$ at lower frequencies. Unfortunately, the inferred Faraday width for FRB 20180916B is much too small to be reliably observed in the corresponding FDFs of bursts detected by CHIME/FRB. However, such an imprint may be observed at lower frequencies where resolution in $\phi$ space is higher. Importantly, the internal depolarization model applied here predicts non-monotonic behavior in $L/I$ at frequencies below 100 MHz for this source. This feature may be leveraged in the future to discriminate between internal and external depolarization effects.

### 4.3. *RM Structure Function and Depolarization*

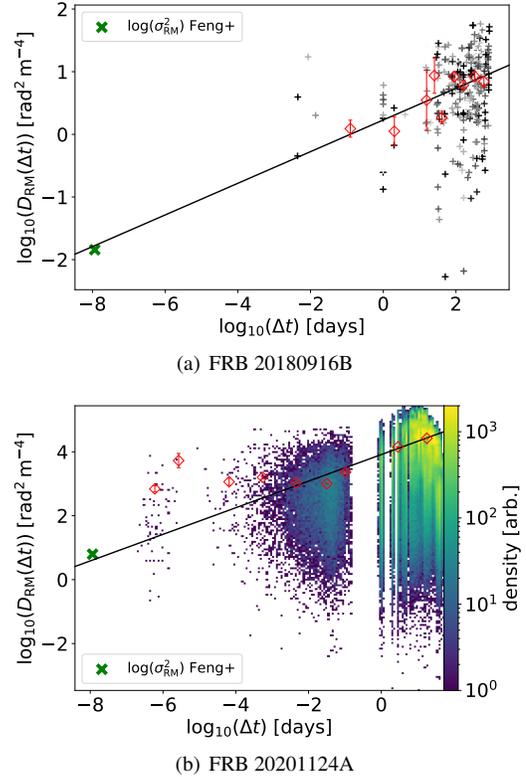

**Figure 7**: The RM structure function, $\log_{10}(D_{\text{RM}}(\Delta t))$, of two prolific repeating FRB sources: FRB 20180916B (stochastic & secular regimes combined) and FRB 20201124A. Panel a) $\log_{10}(D_{\text{RM}})$ measurements are indicated by gray crosses with a color scale saturating on more significant measurements relative to their errors. Panel b) $\log_{10}(D_{\text{RM}})$ measurements are indicated by a two-dimensional histogram with a color scale indicating the density of measurements per bin. Red markers indicate rebinned data and correspond to the mean values with uncertainties estimated as the standard error on the mean. The green marker in each panel corresponds to the $\log(\sigma_{RM}^2)$ constraint determined for FRB 20180916B ($\sigma_{\text{RM}} = 0.12 \ \text{rad m}^{-2}$) and FRB 20201124A ($\sigma_{\text{RM}} = 2.5 \ \text{rad m}^{-2}$; Feng et al. 2022) where a characteristic timescale of $\Delta t = 1$ ms has been chosen. Power-law fits to $D_{\text{RM}}(\Delta t)$ are indicated by black lines and are extrapolated down to timescales probed by $\sigma_{\text{RM}}$ measurements.

Magneto-ionic structures producing burst-to-burst RM variations can only be studied down to physical scales that are determined by the minimum separation between bursts. For PSR J1745−2900, a high cadence of burst detections and additional information that included the source distance and proper motion enabled an estimate of the smallest scale at which magneto-ionic structure produced significant RM variability (Desvignes et al. 2018). In the case of FRB 20180916B, our 44-burst sample has minimum burst sep-arations that are substantially larger, which limits our ability to study magneto-ionic structures at equivalent scales. However, combining the RM structure function analysis with information obtained from depolarization models may offer a powerful method for probing the magneto-ionic environment of repeaters on a wide range of scales. Here, we leverage recent $\sigma_{\mathrm{RM}}$ estimates of Feng et al. (2022) to probe the magneto-ionic environment at physical scales several order of magnitude smaller than what can be determined from burst-to-burst RM variations. In Figure 7, we show the results of structure function analysis applied to RM measurements obtained here for FRB 20180916B (panel a) as well as those recently reported for FRB 20201124A (panel b; Xu et al. 2021). $\sigma_{RM}$ measurements of each repeater are represented as green 'X' markers. These latter measurements are set at a characteristic timescale of $\Delta t \sim 1$ ms, since RM variability has to occur within the burst duration to cause the observed depolarization. $D_{\mathrm{RM}}(\Delta t)$ distributions, excluding the $\sigma_{\mathrm{RM}}$ measurements, for both sources were fit with a power-law relation ($D_{\mathrm{RM}}(\Delta t) \sim \Delta t^l$) and are relatively flat compared to expectations for a Kolmogorov turbulent medium ($l = 5/3$) or equivalent measurements from some pulsars (e.g., McKee et al. 2018; Desvignes et al. 2018). Although neither distribution is adequately described by a power-law relation, it does capture the general timescale dependence and extrapolating down to $\Delta t \sim 1$ ms results in $D_{\mathrm{RM}}$ values that are fairly close to what is predicted from the $\sigma_{\mathrm{RM}}$ measurements. This suggests that the medium producing the RM variations on longer timescales may be the same medium producing the observed depolarization on much shorter timescales. Similar conclusions have been drawn by Wang et al. (2022) for FRB 20201124A based on the evidence for increased depolarization at times when burst-to-burst RM variability is greatest. Such a result was interpreted as an imprint of orbital phase of a binary system, which curiously may also explain the poor fit of $D_{\mathrm{RM}}(\Delta t)$ with by a power-law since RM variability changes significantly over the timescale of these observations.

### 4.4. *Implications for FRB Emission Models*

Numerous emission models exist in the literature seeking to describe the 16.33 day activity cycle of FRB 20180916B. Many of these models divide into rotational, precessional or orbital frameworks. The rotational model posits the existence of a population of ultra long period magnetars, where some combination of particle winds and fallback accretion promotes magnetic activity beyond typical timescales at which termination of activity would occur (Beniamini et al. 2020). In the precessional framework, the periodic activity is produced by precession of an FRB-emitting magnetar, with the precession being free (Levin et al. 2020; Zanazzi & Lai 2020), forced (Sob'yanin 2020) or induced from an orbit

(Yang & Zou 2020) with a companion. In the orbital model, modulation in the activity is caused by an orbit of the source with a companion star that is either a massive star or a millisecond pulsar (Lyutikov et al. 2020; Ioka & Zhang 2020). Many of these models make implicit if not explicit predictions on the polarized behavior of the source as a function of time.

PA behavior over time is considered a key diagnostic for distinguishing between various precessional/rotational models. The lack of polarization calibration applied to these data (Section 2) prevents absolute PA measurements from being determined. For CHIME/FRB observations, only relative PA behavior can be confidently shown, as in the case of PA curves. However, we find that the distribution of uncalibrated PA values deviates substantially from a uniform distribution expected in the case of uncalibrated data. Figure A.2 demonstrates this, displaying the RM and PA nested samples output from the QU-fitting routine for all 44 bursts from this source. 55% of the nested sample PA values cluster between $110° \leq \mathrm{PA} \leq 160°$. Although the precise value of PA reported here is uninformative, the clustering is qualitatively consistent with recent calibrated polarimetric observations where PA variability is $\lesssim 10° - 20°$ within an active cycle (1.7 GHz; Nimmo et al. 2021a) and $\lesssim 40°$ across active cycles (1.4 GHz; Pastor-Marazuela et al. 2020). Assuming that the PA clustering of Figure A.2 is not a feature of some unknown systematic, these observations provide further support to PA stability across activity cycles and on much longer time intervals than in previous studies (Pastor-Marazuela et al. 2020; Nimmo et al. 2021a). However, these constraints on PA variability still significantly exceed the $\lesssim 10°$ precision required to constrain different configurations of precessional models (Li & Zanazzi 2021).

The significant RM evolution offers additional information about the local magneto-ionic environment that may indirectly constrain models. For so-called binary comb models (Lyutikov et al. 2020; Ioka & Zhang 2020), RM variations from the companion star's wind are expected but are highly sensitive to the assumed magnetic properties of the star and the geometry of the system (Lyutikov et al. 2020). Ioka & Zhang (2020) estimate an upper limit of $\lesssim 20$ rad m$^{-2}$ from a Be star companion and argue that the lack of significant RM variability indicates that excess RM arises at a farther distance. While our observations indicate that RM variations are indeed significant, the relevant time scales are much larger than the 16.33 activity cycle of the source. This strongly indicates that, within the framework of a binary comb model, the medium producing Faraday rotation is local to the source but is spatially distinct from the hypothetical medium that obscures emission and thus modulates burst detection. The secular $|\mathrm{RM}|$ decrease over the 2021 April$-$ 2021 December interval will provide useful input in future models seeking to explain this source's periodic bursting activity.

Finally, in addition to constraints on progenitor models arrived at through RM/PA behavior, it has been argued elsewhere that the $\sim 3 - 4$ $\mu s$ burst structure displayed from this source supports a magnetar model in which emission occurs relatively close to the neutron star (Beniamini & Kumar 2020; Lyutikov et al. 2020; Lu et al. 2020) versus farther out in a relativistic shock (Beloborodov 2017; Margalit & Metzger 2018). The short timescale structure observed here in both total intensity and PA is consistent with that previously reported by Nimmo et al. (2021a) at higher frequencies, implying, at least qualitatively, a similar interpretation for emission in the 400-800 MHz CHIME band. Given the recent establishment of chromatic dependence in the burst properties of this source that includes a decreasing linear polarization fraction at lower frequencies and a burst activity window that becomes both narrower and earlier at higher frequencies (Pastor-Marazuela et al. 2020; Pleunis et al. 2021b), it would not be unreasonable to expect a similar frequency dependence in the burst microstructure. A comparison of the timescale of this micro-temporal structure using the brightest bursts of this sample is left for future study. Any systematic difference in this structure as a function of frequency would greatly inform the nature of the emission region.

### 4.5. *Burst Morphology*

The morphology of bursts (the change in flux as a function of time and frequency) from this source in the CHIME band has been discussed before (CHIME/FRB Collaboration et al. 2019, 2020a) but this paper contains the first larger sample of burst dynamic spectra from CHIME/FRB that are derived from baseband data. Baseband data have higher time resolution and also typically higher S/N (because the data have been beamformed towards the best-known sky position of this source) than intensity data, which only has 0.98304-ms time resolution and is extracted from static beams – with many bursts detected away from the centers of those. Note that the downsampling of data products here has been optimized for the analysis of burst polarimetry. We briefly outline qualitative findings regarding the morphology of the bursts in the sample and we also discuss a morphology-induced bias that affects the interpretation of DMs. A more detailed analysis of the burst morphologies will be presented elsewhere.

Inspection of Figure 1 reveals that bursts from FRB 20180916B show similar characteristics in the CHIME baseband data as were identified at coarser time resolution. The bursts typically show wide burst emission envelopes, of up to $\sim 40$ ms (e.g., burst 5). Some more narrow subbursts with widths down to $\lesssim 80$ $\mu s$ are present as have been detected before in bursts from this source and FRB 20200120E (Nimmo et al. 2021a,b). Many of the bursts studied here show narrowband emission ($\sim$50–200 MHz) with varying central frequency that as-of-yet has no obvious pattern as a function of time[12] (or activity phase). Downward-drifting subbursts (the "sad trombone effect") are ubiquitous and typical drift rates seen here agree with previous observations in the CHIME band, with linear drift rates of a few to a few tens of MHz ms$^{-1}$. Note that the boundary between calling something multiple subbursts under one burst envelope or multiple separate burst envelopes without a "bridge" in emission remains fuzzy (e.g., bursts 1, 2, 3, 5; see discussion in Pleunis et al. 2021a).

Finally, it remains difficult to interpret burst structure in case detections have limited S/N or time resolution and bursts appear as "smudges" (see, e.g., Fig. 7 in Gourdji et al. 2019). If downward-drifting subbursts remain unresolved, any DM optimization (for S/N as well as for burst structure) will measure DMs that are biased high, as the algorithms tend to superimpose the (unresolved) subbursts. DMs measured in high S/N detection where individual sharp features are resolved seem to best represent the "true" DM of the source (see also, e.g., CHIME/FRB Collaboration et al. 2020a). That value can then be applied to other lower S/N bursts that are detected close-by in time. Note though that over multi-year timescales changes in the average DM of a few pc cm$^{-3}$ have been observed (e.g., Oostrum et al. 2020).

## 5. CONCLUSION

Multi-year polarimetric monitoring with CHIME/FRB of periodic FRB source FRB 20180916B has provided a sample of 44 bursts recorded with the instrument's baseband system. The native 2.56 $\mu s$ time resolution of these data, combined with the post hoc capability to beamform towards the source's localized position, implies improved measurements of many properties. In this paper, we focus on studying polarized quantities such as the Faraday rotation measure and linear polarization fraction in addition to a handful of other properties. In general, the morphologies of the bursts studied here are qualitatively similar to characteristics described in previous observations based on data with coarser time resolution (e.g., Pleunis et al. 2021a).

We find that RM evolution of FRB 20180916B can be split into two regimes: an epoch (2018 December−2021 April) in which RM variations are quasi-stochastic and another (2021 April−2021 December) in which they are secular. Over the latter regime, we observe a secular decrease of $\sim 50$ rad m$^{-2}$ in the |RM|, with most recent RMs of $\sim -70$ rad m$^{-2}$ significantly different from previous reported measurements (CHIME/FRB Collaboration et al. 2019;

---

[12] Notwithstanding the possible, but still uncertain, spectral evolution occurring between the stochastic and secular RM regimes of this source (Section 3.4).

[Chawla et al. 2020](#); [Pleunis et al. 2021b](#); [Sand et al. 2021](#)) and now consistent with the $\mathrm{RM_{MW}} = -94 \pm 45 \ \mathrm{rad\,m^{-2}}$ contribution estimated for the Milky Way. This interval is not accompanied by any noticeable DM evolution, which suggests the observed RM evolution is largely a product of a changing magnetic field within the Faraday-active medium. Polarimetric observations over the next year will be crucial for constraining models seeking to explain the remarkably linear RM evolution of this source.

We find a marginally significant difference between RM variability of bursts occurring within the same activity cycle compared to those occurring across cycles. However, the strong modulation of the $\mathrm{RM}$ on long timescales is a significant challenge for accurately probing any hypothetical relationship between RM and activity cycle phase. Studying the residuals of a linear fit for the RM evolution over the secular regime allows us to probe for a hypothetical relationship between RM and phase. Interestingly, bursts occurring in the secular regime appear to be at systematically lower frequencies compared to the stochastic regime. While this observation may be coincidental, it may also suggest a possible relationship between the medium modulating the emission and RM variability; this is potentially useful input for models describing the periodic bursting activity of this source.

Finally, the degree of linear polarization of bursts studied here is generally high and consistent with recent predictions for the low level of depolarization expected for this source if observed over $400-800$ MHz ([Feng et al. 2022](#)). However, a small fraction of the sample displays $L/I$ values significantly below $100\%$, which is unlikely to be an imprint of depolarization and instead reflects $L/I$ variations that are likely intrinsic to the source. Such intrinsic variability in $L/I$ has several pulsar analogues, with single pulse studies demonstrating a positive correlation between the brightness of individual pulses and $L/I$. Whether this property extends to repeating FRBs remains to be seen but should motivate future high sensitivity polarimetric observations of these mysterious sources.

The Dunlap Institute is funded through an endowment established by the David Dunlap family and the University of Toronto. R.M. recognizes support from the Queen Elizabeth II Graduate Scholarship and the Lachlan Gilchrist Fellowship. B.M.G. is supported by an NSERC Discovery Grant (RGPIN-2015-05948), and by the Canada Research Chairs (CRC) program. K.W.M. is supported by an NSF Grant (2008031). V.M.K. holds the Lorne Trottier Chair in Astrophysics & Cosmology, a Distinguished James McGill Professorship, and receives support from an NSERC Discovery grant (RGPIN 228738-13), from an R. Howard Webster Foundation Fellowship from CIFAR, and from the FRQNT CRAQ. A.B.P. is a McGill Space Institute (MSI) Fellow and a Fonds de Recherche du Quebec – Nature et Technologies (FRQNT) postdoctoral fellow. Z.P. is a Dunlap Fellow. C.L. was supported by the U.S. Department of Defense (DoD) through the National Defense Science & Engineering Graduate Fellowship (NDSEG) Program. K.S. is supported by the NSF Graduate Research Fellowship Program. E.P. acknowledges funding from an NWO Veni Fellowship. FRB research at UBC is supported by an NSERC Disocvery Grant and by the Canadian Institute for Advanced Research. The CHIME/FRB baseband system is funded in part by a CFI JELF award to IHS. The polarization analysis presented here makes use of the `RMtools` package[13]([Purcell et al. 2020](#)) written by Cormac Purcell, and maintained by Cameron Van Eck.

---

[13] https://github.com/CIRADA-Tools/RM-Tools

APPENDIX

A. SUPPLEMENTAL FIGURES

B. RM-SYNTHESIS AND QU-FITTING SUMMARY PLOTS

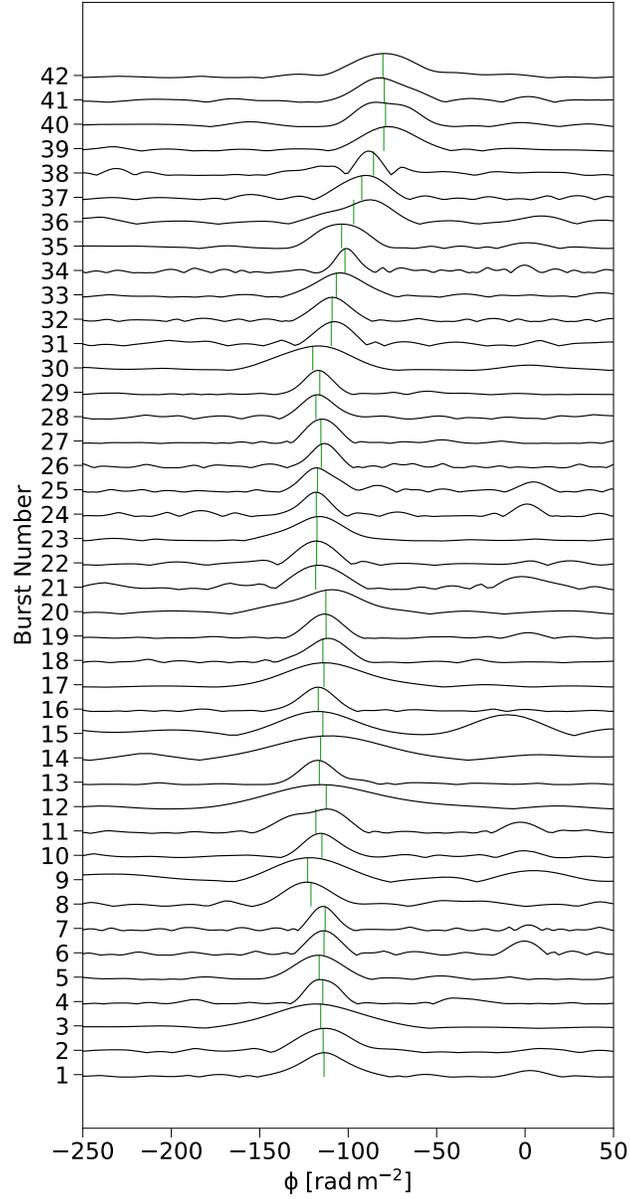

**Figure A.1**: Cleaned FDFs (see Section 2.2.2) for bursts from FRB 20180916B. Each curve is normalized to a unitary peak and ordered chronologically by burst number (see Table 1). $\mathrm{RM_{QU}}$ measurements are indicated as vertical green lines. Peaks near $\phi \sim 0 \,\mathrm{rad\,m^{-2}}$ correspond to instrumental polarization.

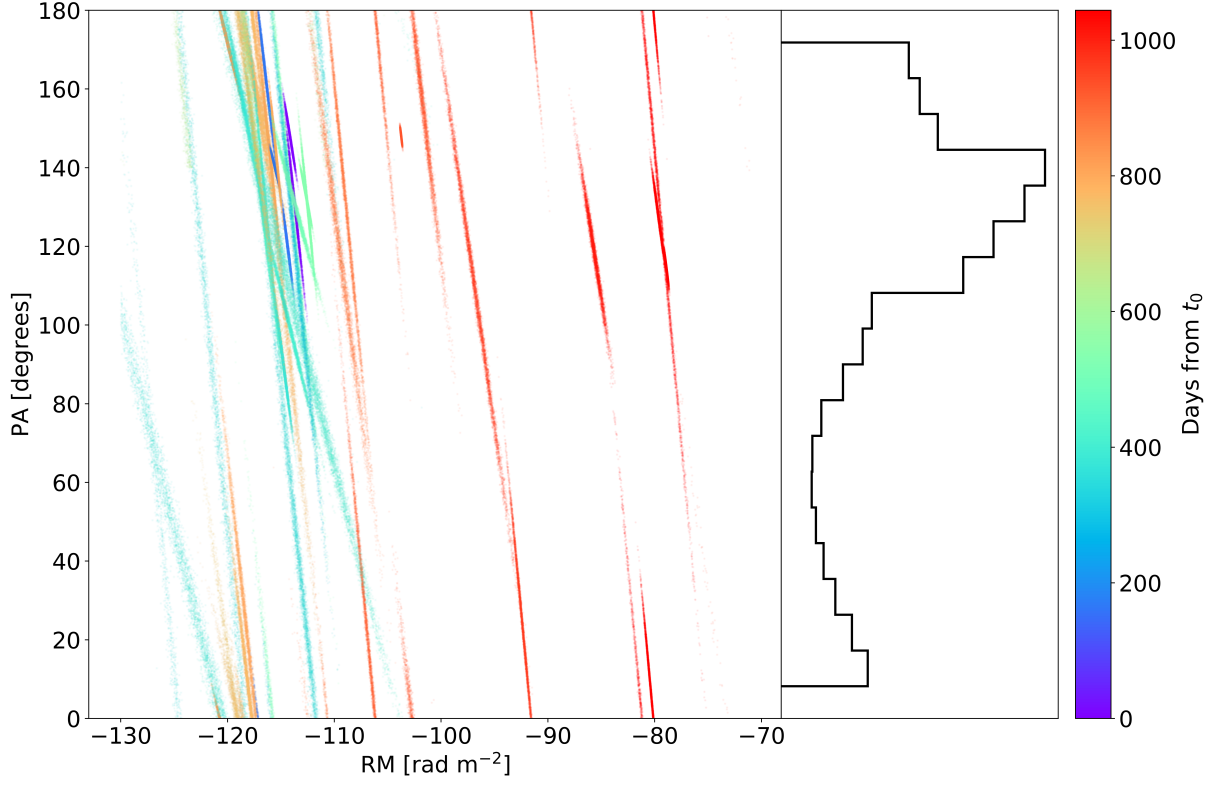

**Figure A.2**: QU-fitting nested samples of RM and the (uncalibrated) PA for each of the 44 bursts from repeating source FRB 20180916B. Color map indicates days since the first burst ($t_0$) at MJD = 58477.16185100. The adjoining panel displays the distribution of PA samples that is significantly different from a uniform distribution expected for uncalibrated polarization observations over a multi-year time interval. Rather, the clustering shown here is consistent with the relative PA stability reported elsewhere for this source (Pastor-Marazuela et al. 2020; Nimmo et al. 2021a).

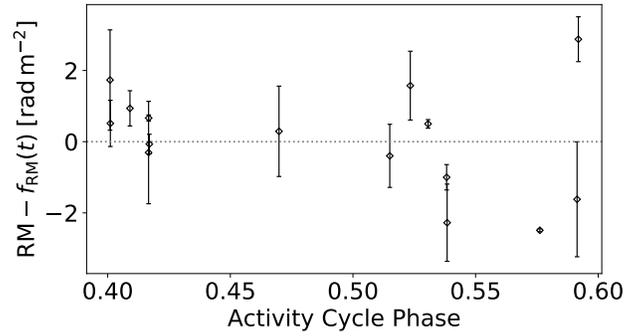

**Figure A.3**: The residual RM versus phase in the 16.33 day activity cycle of FRB 20180916B for 14 bursts occurring during the secular regime (2021 April−2021 December). RM residuals (RM − $f_{RM}(t)$) are determined by subtracting values obtained from the linear fit of Figure 2.

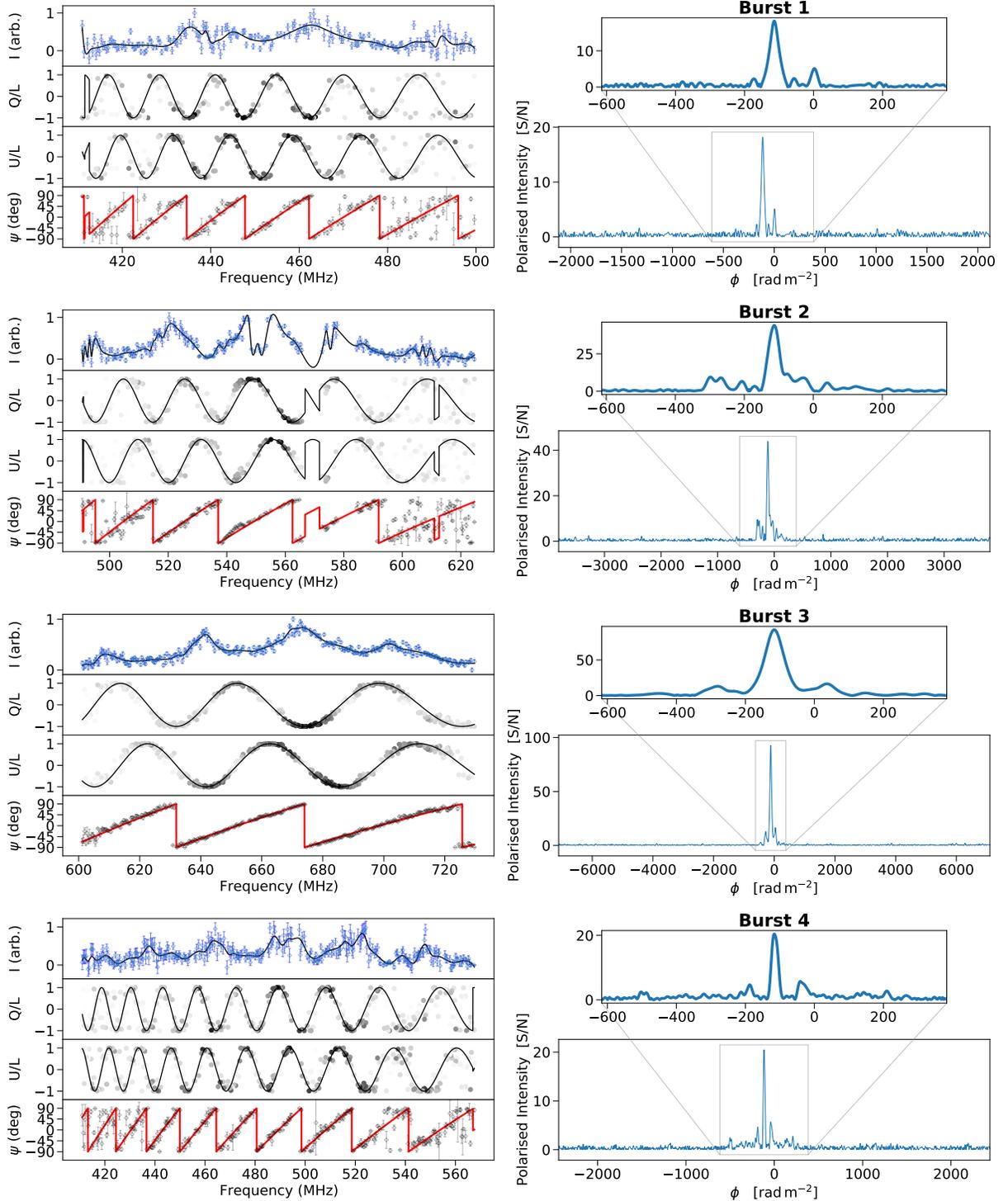

**Figure B.1**: The results of QU-fitting (left) and RM-synthesis (right) for individual bursts from repeating source FRB20180916B after correcting for a non-zero $\tau$ parameter (see Section 2.2.3). Left panel: The polarization spectra displaying, from top to bottom, Stokes $I$ (blue points) and its cubic spline smoothed version (black line), Stokes $Q$ parameter divided by the total linear polarization ($L$), Stokes $U$ divided by the total linear polarization, and the uncalibrated polarization angle ($\psi_0$). Frequency channels with highly polarized signal are indicated with darker points. Best-fit models are indicated as black lines (red lines for $\psi_0$). Right panel: The cleaned FDFs displaying linear polarized intensity as a function of $\phi$ (RM). The upper panel shows the FDFs confined to $\pm 500\,\mathrm{rad\,m^{-2}}$ of their peak locations.

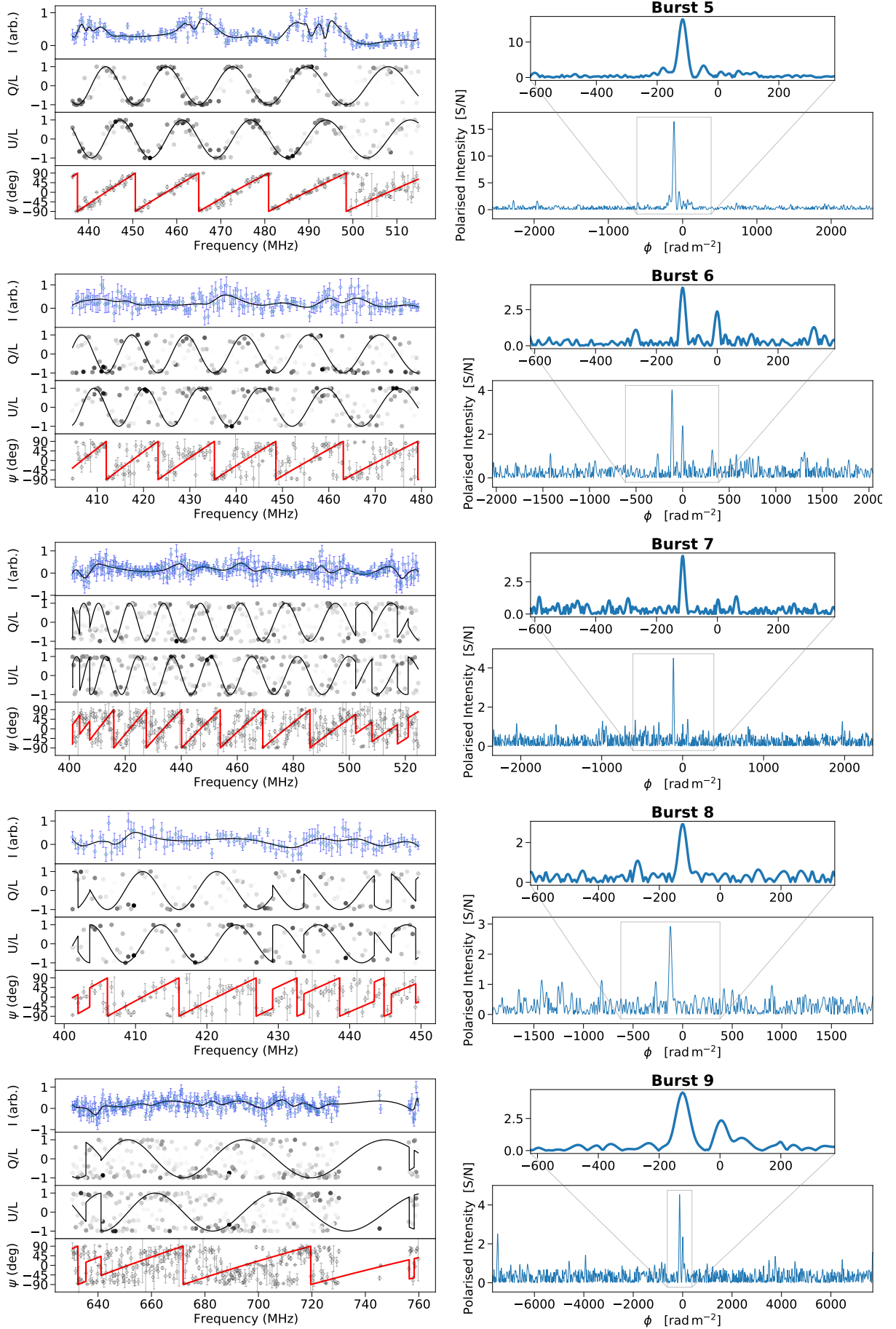

**Figure B.1**: (Continued) The results of QU-fitting (left) and RM-synthesis (right) for individual bursts from repeating source FRB20180916B.

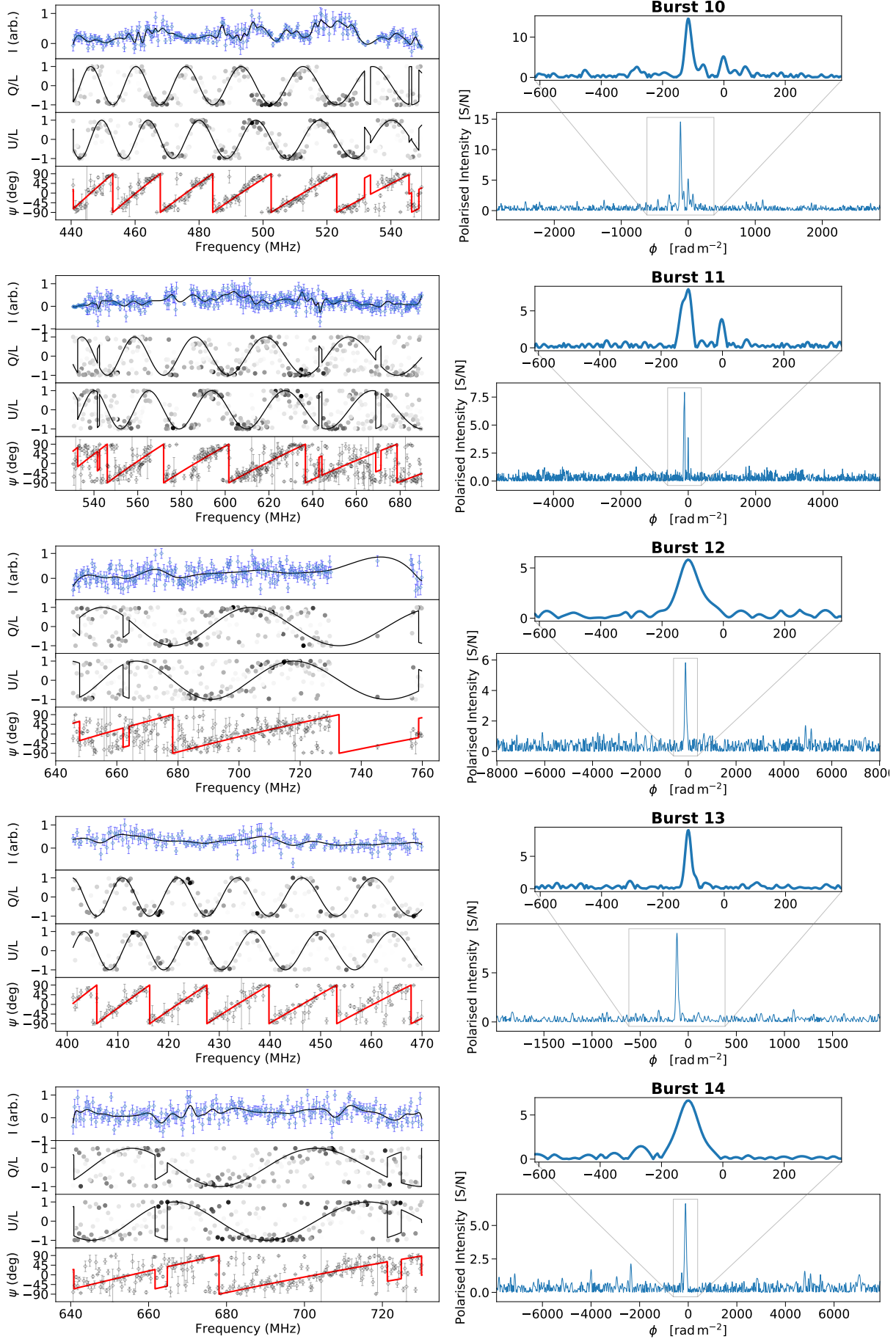

**Figure B.1**: (Continued) The results of QU-fitting (left) and RM-synthesis (right) for individual bursts from repeating source FRB20180916B.

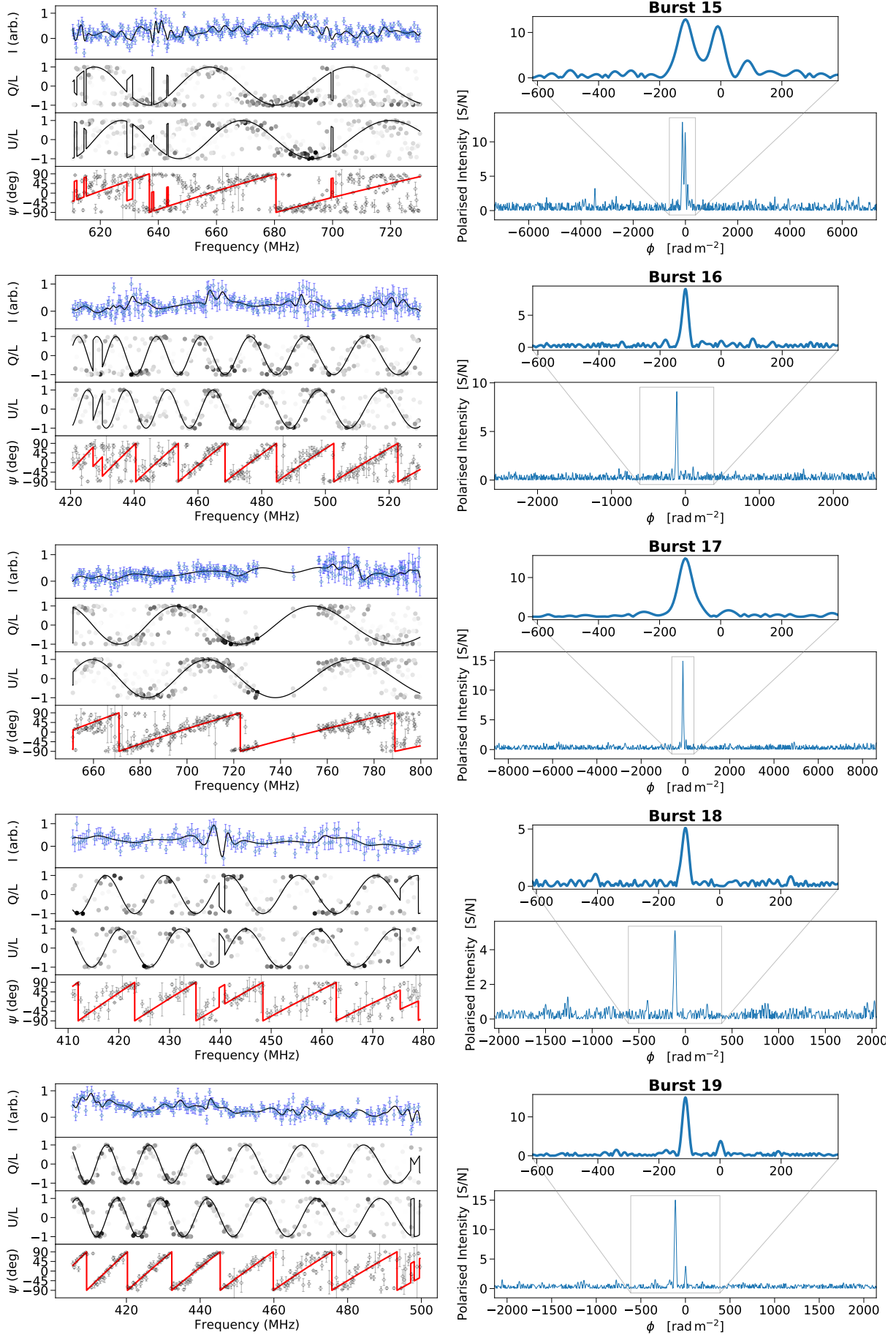

**Figure B.1**: (Continued) The results of QU-fitting (left) and RM-synthesis (right) for individual bursts from repeating source FRB20180916B.

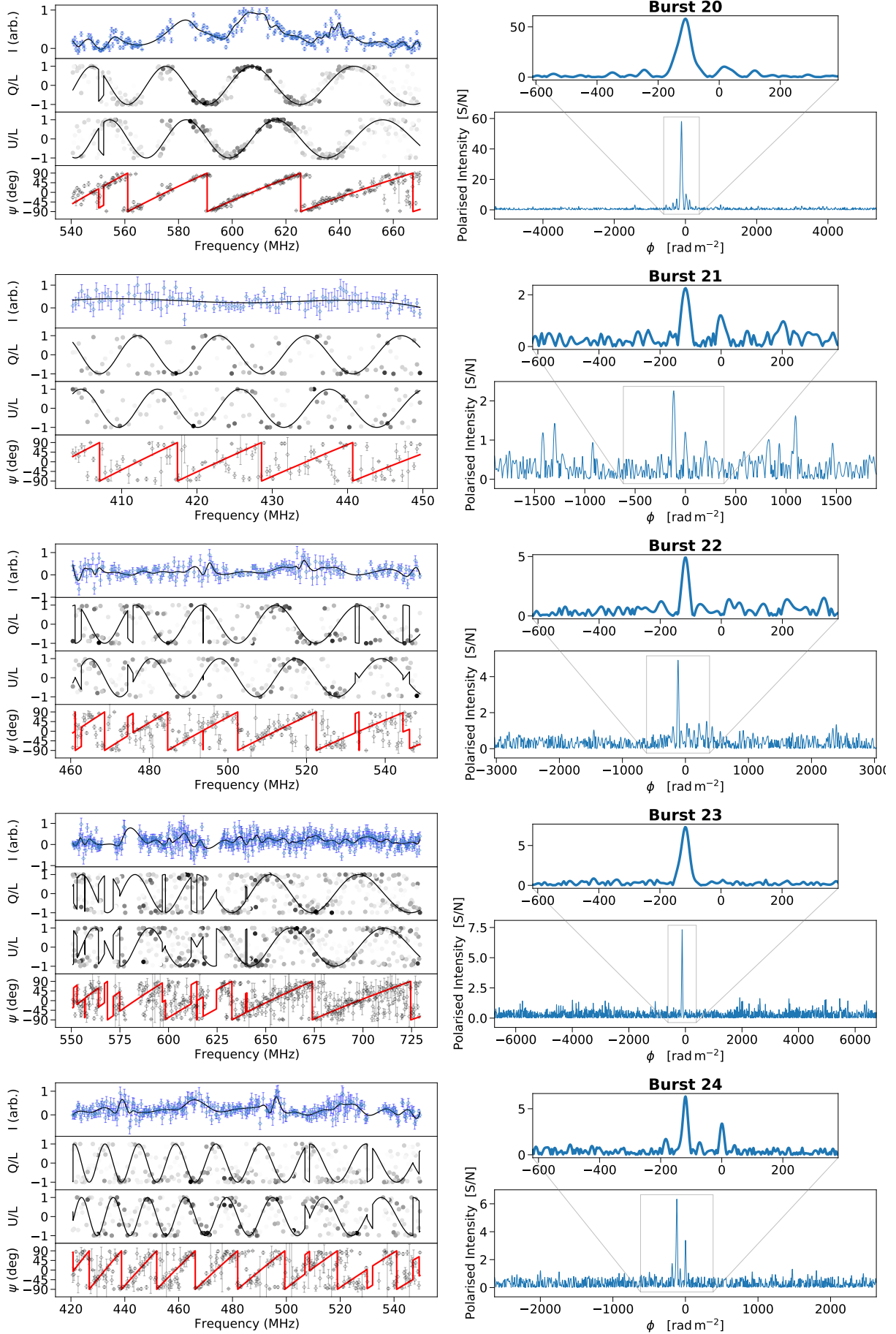

**Figure B.1**: (Continued) The results of QU-fitting (left) and RM-synthesis (right) for individual bursts from repeating source FRB20180916B.

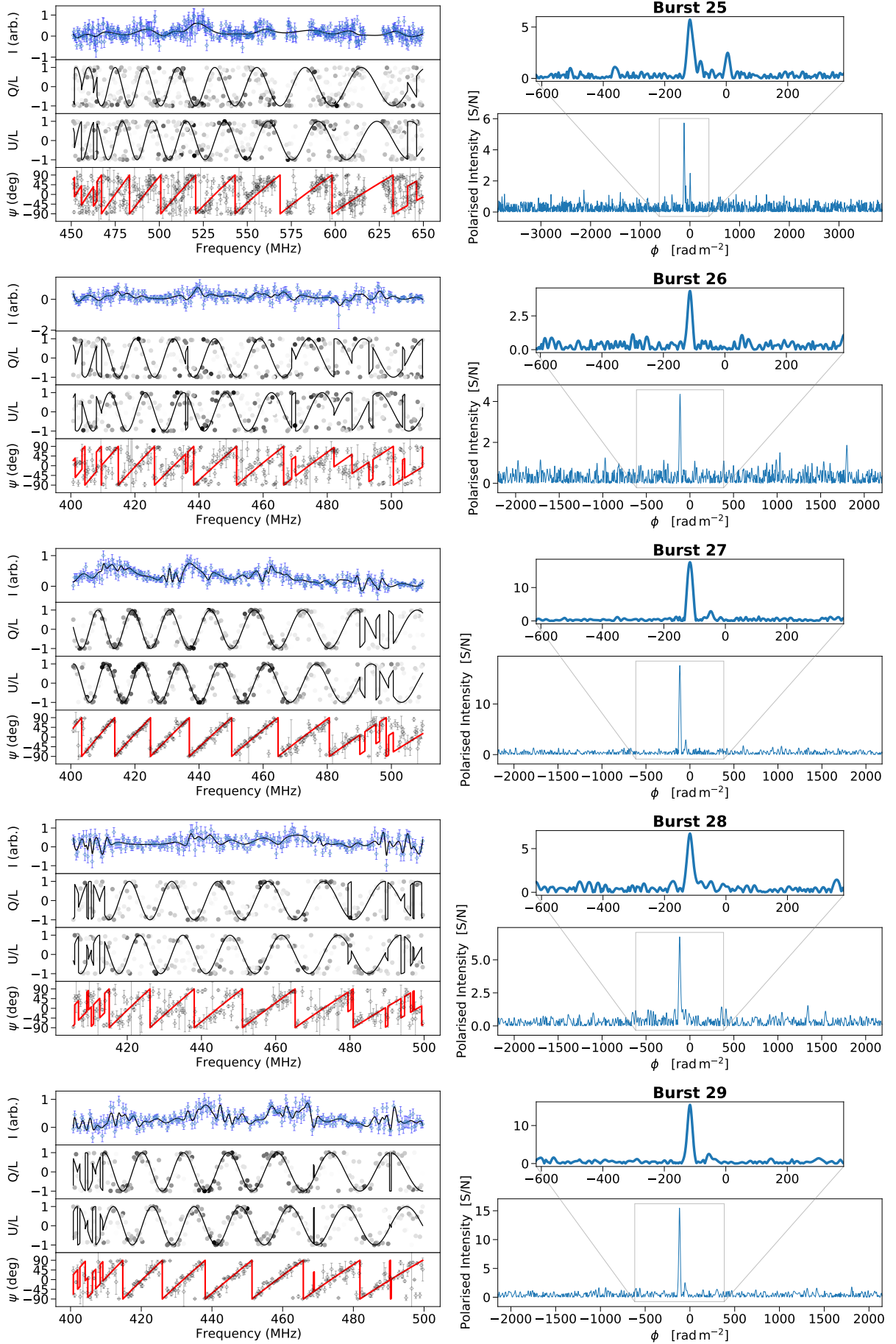

**Figure B.1**: (Continued) The results of QU-fitting (left) and RM-synthesis (right) for individual bursts from repeating source FRB20180916B.

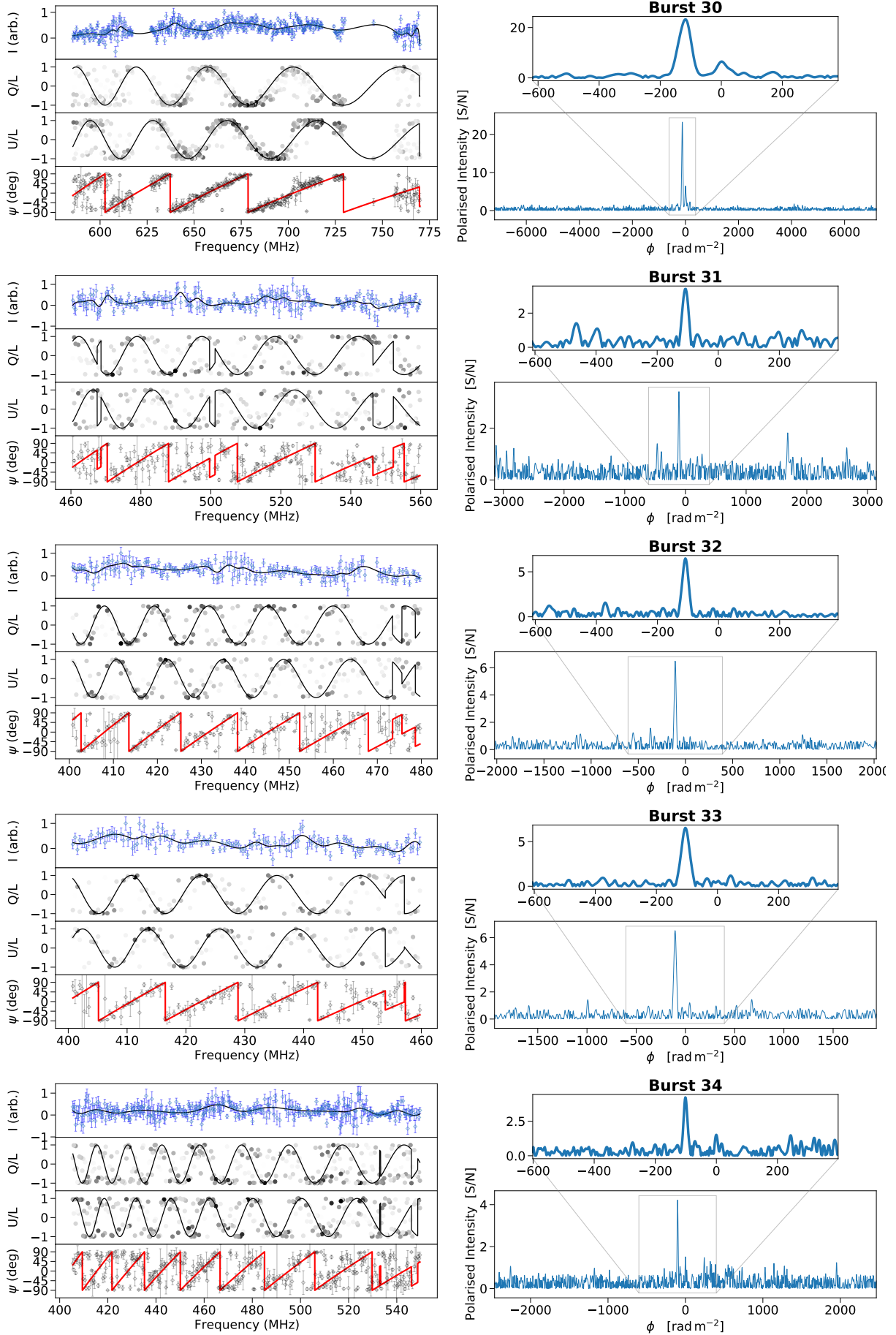

**Figure B.1**: (Continued) The results of QU-fitting (left) and RM-synthesis (right) for individual bursts from repeating source FRB20180916B.

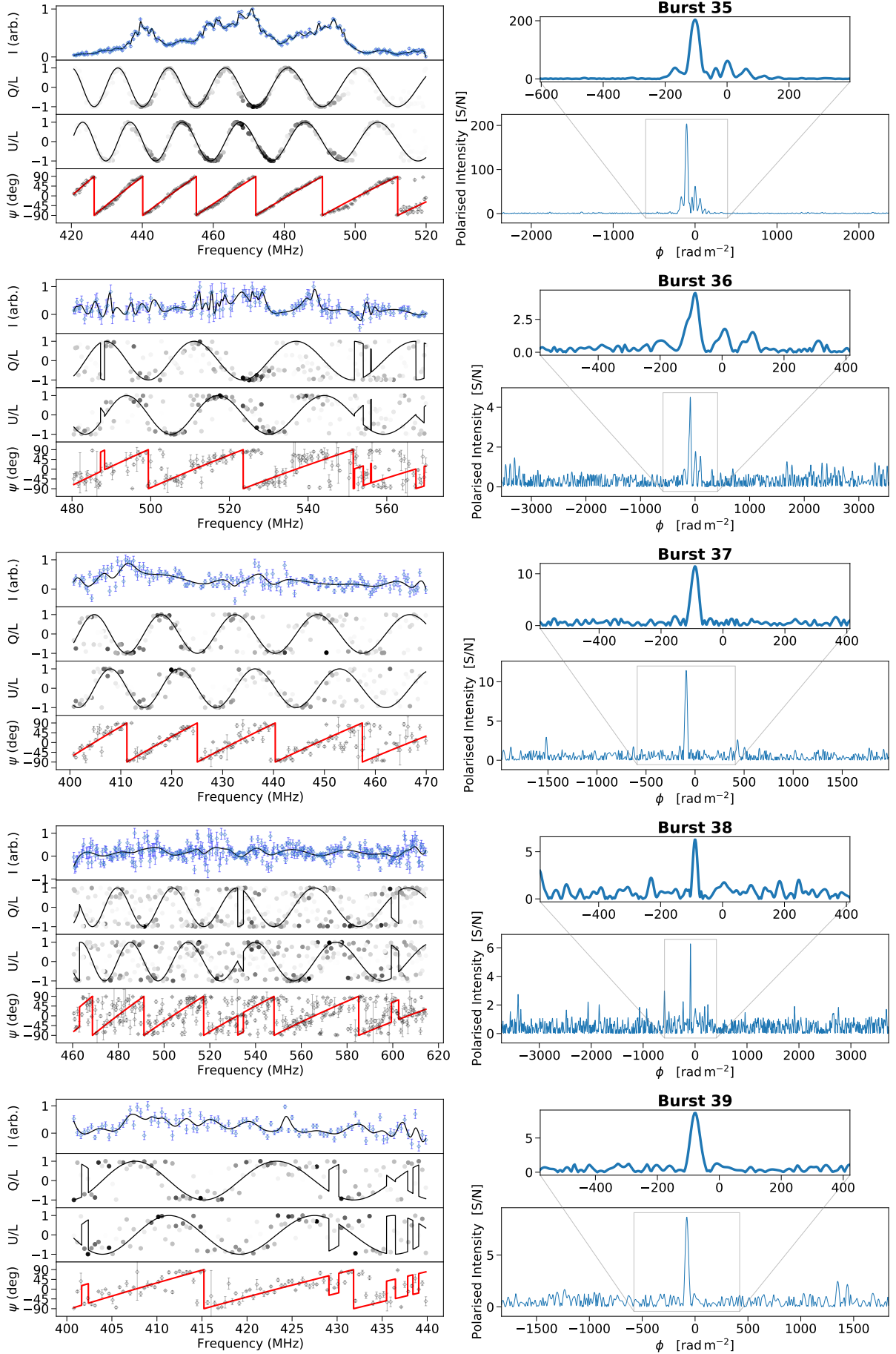

**Figure B.1**: (Continued) The results of QU-fitting (left) and RM-synthesis (right) for individual bursts from repeating source FRB20180916B.

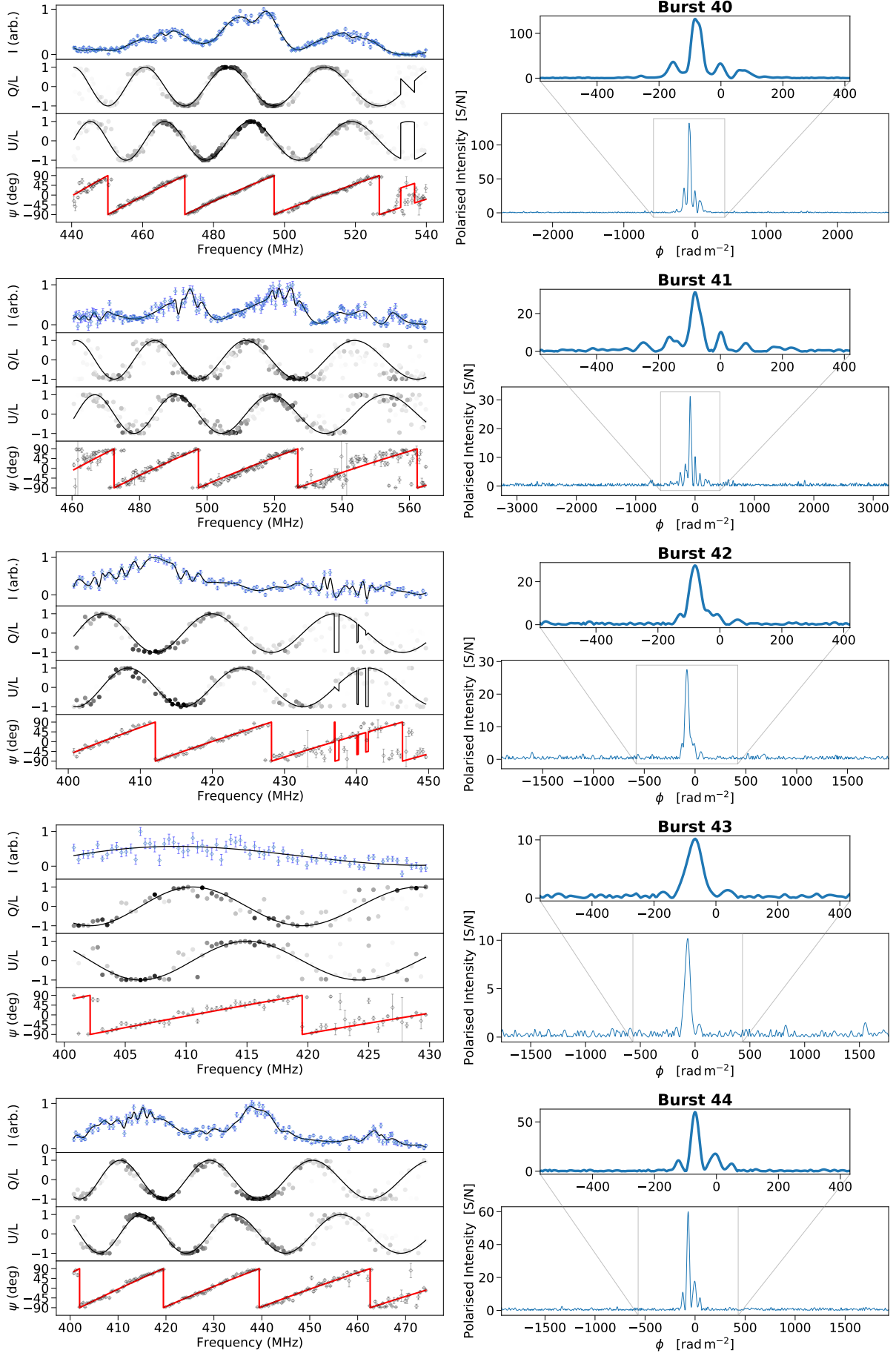

**Figure B.1**: (Continued) The results of QU-fitting (left) and RM-synthesis (right) for individual bursts from repeating source FRB20180916B.